\documentclass[journal]{new-aiaa}
\usepackage[utf8]{inputenc}
\usepackage{textcomp}

\usepackage{graphicx}
\usepackage{amsmath}
\usepackage[subrefformat=parens]{subcaption}
\usepackage{bm}
\usepackage[version=4]{mhchem}
\usepackage{siunitx}
\usepackage{longtable,tabularx}
\usepackage{algorithmic}
\usepackage{algorithm}
\usepackage{tikz}
\usepackage{placeins}

\hypersetup{hidelinks}

\newcommand{\Secref}[1]{Section~\ref{#1}}
\newcommand{\Eqref}[1]{Eq.~(\ref{#1})}
\newcommand{\Eqsref}[1]{Eqs.~(\ref{#1})}

\newcommand{\Figref}[1]{Fig.~\ref{#1}}
\newcommand{\Tabref}[1]{Table~\ref{#1}}

\newcommand{\argmin}{\mathop{\rm arg~min}\limits}

\usetikzlibrary{positioning,calc,fit,backgrounds,arrows.meta}

\begin{document}

\title{Stochastic Differential Dynamic Programming for \\ Trajectory Optimization under Partial Observability}

\begingroup
\renewcommand{\thefootnote}{}
\renewcommand{\theHfootnote}{conferencepaper.1}
\footnotetext{An earlier version of this paper was presented as AAS Paper 24-424 at the 2024 AAS/AIAA Astrodynamics Specialist Conference, Broomfield, CO, August 11--15, 2024.}
\endgroup

\author{
Masahiro Fujiwara\footnote{Researcher, Research and Development Directorate, Japan Aerospace Exploration Agency, Sagamihara, Kanagawa 252-5210, Japan; Member AIAA; Corresponding author: fujiwara.masahiro@jaxa.jp.}
\ and
Naoya Ozaki\footnote{Associate Professor, Institute of Space and Astronautical Science, Japan Aerospace Exploration Agency, Sagamihara, Kanagawa 252-5210, Japan; Member AIAA.}\\
{\normalsize\itshape Japan Aerospace Exploration Agency, Sagamihara, Kanagawa 252-5210, Japan}
}

\maketitle{}

\begin{abstract}
Designing spacecraft trajectories remains challenging in the presence of stochastic effects such as maneuver execution errors and observation uncertainties. Although covariance control and belief-space planning provide useful tools for designing robust control policies and information-aware trajectories under uncertainty, practical methods remain limited for partially observable trajectory optimization problems in which trajectory design, orbit determination, and correction maneuver planning are tightly coupled. This paper presents a stochastic differential dynamic programming algorithm for such coupled problems. The proposed method optimizes the nominal control sequence and feedback gains subject to a belief-state transition model and general mission constraints, explicitly accounting for the dependence of covariance propagation on the nominal trajectory without relying on the separation principle. Numerical examples demonstrate that the proposed algorithm produces navigation-aware and uncertainty-robust solutions across a range of dynamical systems, observation models, and uncertainty levels.
\end{abstract}

\section*{Nomenclature}

{\renewcommand\arraystretch{1.0}
\noindent\begin{longtable*}{@{}l @{\quad=\quad} l@{}}
$\bm{0}_n$ & $n$-dimensional zero vector \\
$I_n$ & $n \times n$ identity matrix \\
$\mathcal{N}\left(\bm{m}, P\right)$ & multivariate Gaussian distribution with mean $\bm{m}$ and covariance $P$ \\
$n_u$ & control dimension \\
$n_w$ & dynamical noise dimension \\
$n_x$ & state dimension \\
$n_y$ & observation dimension \\
$O_n$ & $n \times n$ zero matrix \\
$O_{n \times m}$ & $n \times m$ zero matrix \\
$p(\cdot)$ & probability density function \\
$\mathbb{R}^n$ & $n$-dimensional real vector space \\
$\mathrm{Cov}\left[\cdot\right]$ & covariance operator \\
$\mathbb{E}\left[\cdot\right]$ & expectation operator \\
$\mathbb{P}\left[\cdot\right]$ & probability \\
$\mathrm{tr}\left(\cdot\right)$ & trace operator \\
$\mathrm{vec}\left(\cdot\right)$ & vectorization operator \\
\end{longtable*}}

\section{Introduction}
Spacecraft trajectory design under uncertainty becomes particularly challenging when the orbit-determination process depends on the nominal trajectory.
In such problems, maneuver design and navigation design cannot be treated independently, because the trajectory determines the measurement geometry and the achievable navigation accuracy.
This coupling is especially important in scenarios with weak observability, such as angles-only navigation or missions with limited tracking opportunities.
Although high-quality measurements are often available in conventional missions, for example, through radiometric tracking with Delta-DOR in deep-space missions~\cite{CurkendallBorder2013DeltaDOR} or GNSS measurements in LEO missions~\cite{AllahvirdiZadehWangElMowafy2022ASCE}, increasing mission complexity and operational cadence can limit such tracking opportunities.
These considerations motivate a unified framework for partially observable trajectory optimization problems, in which maneuver design, orbit determination, and correction maneuver planning are tightly coupled under uncertainty.

In spaceflight applications, covariance control approaches~\cite{HotzSkelton1987CovarianceControl} have been developed to design maneuver policies under uncertainty using a variety of optimization techniques, including differential dynamic programming (DDP), nonlinear programming (NLP), and sequential convex programming (SCP).
Refs.~\cite{Ozaki2018-rx, Ozaki2020-id} derive tube stochastic DDP, in which DDP is applied to a stochastic dynamical system sampled by the unscented transform.
As a higher-fidelity approach, Ref.~\cite{Greco2022-at} presents an NLP-based method for optimal impulsive control in belief space, incorporating the orbit-determination process into uncertainty propagation via polynomial chaos expansion.
Ref.~\cite{Varghese2026-sh} improves convergence of NLP with covariance dynamics by exploiting the forward--backward structure and introducing a feedback-gain parameterization that reduces the search space.
SCP-based methods have also been extensively studied due to their tractability and the availability of efficient convex optimization solvers.
For example, Ref.~\cite{ridderhof2020chance} formulates convexified discrete-time dynamics, costs, and constraints that can be solved iteratively even for nonlinear trajectory optimization, and Ref.~\cite{Oguri_undated-kc} incorporates stochastic mass dynamics and the orbit-determination process into an SCP framework.
Because the feedback gain is parameterized as a block lower-triangular matrix that depends on past states at the discretization nodes in many of these approaches, the computational effort increases quadratically with the number of nodes.
To improve the efficiency of covariance control with chance constraints, sequential semidefinite programming methods have also been proposed~\cite{rapakoulias2023discretetimeoptimalcovariancesteering, pilipovsky2024computationallyefficientchanceconstrained} and applied to astrodynamics problems in Ref.~\cite{Kumagai2025-bl}.

Planning under dynamical and observation uncertainties has also been studied extensively in the robotics community, where the problem is often formulated as a partially observable Markov decision process and addressed via belief-space planning~\cite{KaelblingLittmanCassandra1998POMDP}.
Because belief-space planning is generally intractable due to its infinite-dimensional state space, practical methods often assume Gaussian beliefs and perform local approximations around a nominal trajectory.
Ref.~\cite{Platt2010-jq} applies a linear quadratic regulator (LQR) to deterministic Gaussian belief-space dynamics under the assumption of maximum-likelihood observation (MLO).
Refs.~\cite{Van_Den_Berg2012-fn, Van_den_Berg2017-br} remove the MLO assumption and propose iterative local optimization methods in belief space by expanding the cost function and dynamics around the nominal trajectory to obtain time-varying affine feedback policies.
Ref.~\cite{Indelman2015-jb} further extends the framework by introducing random binary variables to model missed observations.
Related active-sensing trajectory generation methods~\cite{8460476, 8827581} have also been developed to improve estimation performance by shaping the nominal trajectory according to information-related criteria. These approaches are closely related to belief-space planning in that they exploit the coupling between motion and estimation.

These two lines of research offer complementary strengths.
Covariance control approaches provide powerful tools for robust control policy design under uncertainty, whereas belief-space planning methods are effective for information-aware decision-making.
However, partially observable trajectory optimization problems in which trajectory design, orbit determination, and correction maneuver planning must be addressed in a unified manner remain insufficiently studied.
Many covariance control formulations simplify the coupling by assuming that the orbit-determination process can be separated from the trajectory optimization problem.
As a result, they are not primarily designed to generate navigation-aware solutions in which the nominal trajectory is deliberately shaped to pass through information-rich regions.
In contrast, belief-space planning methods primarily emphasize information gathering through nominal-trajectory design and generally do not explicitly optimize feedback policies under general mission constraints.
Thus, the joint optimization of informative nominal trajectories and feedback policies under general mission constraints remains a challenging problem in spacecraft mission design.

To address this gap, we develop a partially observable stochastic differential dynamic programming (PO-SDDP) framework for partially observable trajectory optimization problems.
The proposed framework enables simultaneous optimization of the nominal control and feedback gains while explicitly accounting for state estimation and covariance propagation under general mission constraints.
The main technical contributions are as follows.
First, we formulate a generalized belief-space DDP framework by augmenting the belief-space transition model with the state-estimate covariance and by treating the feedback gain as an optimization variable.
Second, we develop a practical formulation for spacecraft mission design by combining an augmented Lagrangian method with regularized or smooth approximations of representative cost functions and constraints, including thrust magnitude and guidance accuracy constraints.
Third, we develop a semi-analytic method to efficiently compute the state transition matrices for covariance propagation, together with automatic differentiation for model-dependent derivatives and model-independent tensor operations for covariance propagation.

From an algorithmic perspective, the proposed method is rooted in stochastic differential dynamic programming~\cite{Theodorou2010-xy} and related DDP-based covariance control methods for nonlinear stochastic systems~\cite{Yi2020-mz}.
The main distinction is that the proposed method formulates the local optimization problem in belief space, explicitly accounting for the dependence of navigation performance on the nominal trajectory.
An observability-aware DDP approach~\cite{fujiwara2024} is also closely related to the present work.
Compared with that approach, the proposed method can be interpreted as a stochastic extension that incorporates covariance propagation and feedback-gain optimization into the DDP recursion.
An earlier version of this work appeared in Ref.~\cite{fujiwara}.
This manuscript extends the conference paper by providing a more rigorous derivation of the dynamics, additional implementation details, and an expanded numerical analysis.

We demonstrate the proposed algorithm through three numerical examples: the light-dark domain problem, an Earth-to-Mars planar transfer, and a periodic-orbit transfer in the Earth--Moon circular restricted three-body problem (CR3BP).
These examples show that the method can address a broad range of problems, from information-aware trajectory shaping to robust correction maneuver design.

The remainder of this paper is organized as follows.
\Secref{sec:belief_propagation} derives a belief-space transition model defined by the dynamics, observation model, and uncertainty model.
In \Secref{sec:po_sddp}, we introduce the partially observable stochastic DDP algorithm.
\Secref{sec:impl_details} presents semi-analytic computations of the state transition matrices for covariance propagation, together with representative cost functions and constraints applicable to space mission design using smoothed approximations.
\Secref{sec:numerical_examples} presents the numerical results, and \Secref{sec:conclusion} concludes the paper.

\section{Belief Propagation for Partially Observable Trajectory Optimization Problems} \label{sec:belief_propagation}
This section derives a tractable belief-space transition model for partially observable trajectory optimization.
In belief-space planning, decisions are made based on a belief over the state rather than the unobserved true state.
Our objective is to construct belief-space dynamics for mission design when future observations are unknown.
To this end, we approximate the belief using up to second-order moments and derive propagation equations for two covariance matrices: the estimation-error covariance and the state-estimate covariance.
These covariance dynamics, together with the nominal dynamics, constitute the belief-space transition model used in the subsequent DDP algorithm.

\subsection{Dynamics, observations, and time discretization}

We consider a controlled stochastic system over the mission design interval $t \in [t_0, t_f]$, with state $\bm{x}(t) \in \mathbb{R}^{n_x}$ and control $\bm{u}(t) \in \mathbb{R}^{n_u}$.
The true state evolves according to the continuous-time stochastic dynamics
\begin{align}\label{eq:stochastic_dynamics}
    \mathrm{d}\bm{x} = \bm{f}(\bm{x}, \bm{u}, t)\mathrm{d}t + G_x(\bm{x}, \bm{u})\mathrm{d}\bm{w}_x
\end{align}
where $\bm{f}: \mathbb{R}^{n_x} \times \mathbb{R}^{n_u} \times \mathbb{R} \rightarrow \mathbb{R}^{n_x}$ represents the deterministic part of the dynamics, $\mathrm{d}\bm{w}_x \in \mathbb{R}^{n_w}$ is a Brownian-motion increment, and $G_x(\bm{x}, \bm{u}): \mathbb{R}^{n_x} \times \mathbb{R}^{n_u} \rightarrow \mathbb{R}^{n_x \times n_w}$ is a weighting matrix that determines the process-noise intensity as a function of the state and control, for example, to model maneuver errors that depend on thrust magnitude and direction.

To discretize the stochastic dynamics in \Eqref{eq:stochastic_dynamics}, the optimization interval $[t_0,t_f]$ is partitioned by maneuver and observation epochs.
Let $N_j(k)$ denote the number of observation epochs between the $k$-th and $(k+1)$-th maneuver epochs.
When $N_j(k)>0$, these observation epochs are ordered as
\begin{align} \label{eq:timeline}
    t_k < t_{k,1} < \dots < t_{k,N_j(k)} \leq t_{k+1}, & & k = 0,\ldots,N-1
\end{align}
where epochs with a single subscript, except for the terminal epoch $t_N=t_f$, correspond to maneuver epochs, whereas epochs with double subscripts correspond to observation epochs.
Without loss of generality, we define $t_{k,0} := t_k$ and $t_{k,N_j(k)+1} := t_{k+1}$.
Thus, although $N_j(k)$ denotes the number of observation epochs, the interval between two consecutive maneuver epochs contains $N_j(k)+1$ intermediate transitions.
Here, the final observation epoch between two consecutive maneuver epochs is allowed to coincide with the latter maneuver epoch.
In such a case, the observation is obtained just before performing the maneuver using the latest orbit-determination result, and therefore $t_{k,N_j(k)} = t_{k,N_j(k)+1} = t_{k+1}$.
Moreover, different observation sources may be available at different observation epochs.
This partitioning is applicable to general space mission design problems.

Observations at each epoch are modeled as
\begin{align}
\bm{y}_{k,j} = \bm{h}_{k,j}(\bm{x}_{k,j}) + G_{y_{k,j}}(\bm{x}_{k,j})\bm{w}_{y_{k,j}}
\label{eq:obs_eq}
\end{align}
where $\bm{h}_{k,j}: \mathbb{R}^{n_x} \rightarrow \mathbb{R}^{n_{y_{k,j}}}$ is the observation model at epoch $t_{k,j}$, $\bm{w}_{y_{k,j}} \in \mathbb{R}^{n_{y_{k,j}}}$ is an independent Gaussian noise vector, and $G_{y_{k,j}}: \mathbb{R}^{n_x} \rightarrow \mathbb{R}^{n_{y_{k,j}} \times n_{y_{k,j}}}$ is a weighting matrix for state-dependent observation noise, with covariance
\begin{align}
W^{-1}_{k,j}(\bm{x}_{k,j}) := G_{y_{k,j}}(\bm{x}_{k,j})G_{y_{k,j}}(\bm{x}_{k,j})^\top
\label{eq:setup_W}
\end{align}
Here, $W_{k,j}(\bm{x}_{k,j})$ is defined as the inverse matrix of the observation noise covariance.
We assume that $\bm{h}_{k,j}$ and $G_{y_{k,j}}$ may vary with the epoch $t_{k,j}$ to enable handling different observation types and dimensions across epochs, and that $\bm{w}_{y_{k,j}}$ is mutually independent across epochs and independent of the process noise $\bm{w}_x$.

\subsection{Belief representation}

Given the past nominal control sequence $\bar{\bm{u}}_{0:k} = \{\bar{\bm{u}}_{0}, \ldots, \bar{\bm{u}}_{k}\}$ and the observation histories $\bm{y}_{k,1:j} = \{\bm{y}_{k,1}, \ldots, \bm{y}_{k,j}\}$ and $\bm{Y}_{k-1} = \{\bm{Y}_{0}, \ldots, \bm{Y}_{k-1}\}$, where $\bm{Y}_{j} = \{\bm{y}_{j,1}, \ldots, \bm{y}_{j,N_j(j)}\}$, the belief is defined as the conditional distribution of the state:
\begin{align}
    \bm{b}(\bm{x}_{k,j}) = p(\bm{x}_{k,j} \mid \bar{\bm{u}}_{0:k}, \bm{Y}_{k-1}, \bm{y}_{k,1:j})
\end{align}
Given a new observation $\bm{y}_{k,j+1}$, the belief is propagated by Bayesian filtering:
\begin{align}\label{eq:Bayesian_filtering}
\begin{aligned}
    \bm{b}(\bm{x}_{k,j+1})
    &= \eta_{k,j+1}\, p(\bm{y}_{k,j+1} \mid \bm{x}_{k,j+1}, \bar{\bm{u}}_{0:k}, \bm{Y}_{k-1}, \bm{y}_{k,1:j})
    \int p(\bm{x}_{k,j+1} \mid \bm{x}_{k,j}, \bar{\bm{u}}_{k})\, \bm{b}(\bm{x}_{k,j})\, \mathrm{d}\bm{x}_{k,j}
\end{aligned}
\end{align}
where $\eta_{k,j+1}$ is a normalizing constant independent of $\bm{x}_{k,j+1}$.
In general, \Eqref{eq:Bayesian_filtering} yields an infinite-dimensional probability distribution that does not admit a closed-form representation.
To address this issue, we first define the state estimate and the estimation-error covariance as
\begin{subequations}\label{eq:belief_moments}
\begin{align}
    \hat{\bm{x}}_{k,j} &:= \mathbb{E}\!\left[\bm{x}_{k,j} \mid \bar{\bm{u}}_{0:k}, \bm{Y}_{k-1}, \bm{y}_{k,1:j}\right]
    \label{eq:state_estimate}\\
    \tilde{P}_{k,j} &:= \mathbb{E}\!\left[(\bm{x}_{k,j} - \hat{\bm{x}}_{k,j})(\bm{x}_{k,j} - \hat{\bm{x}}_{k,j})^\top \mid \bar{\bm{u}}_{0:k}, \bm{Y}_{k-1}, \bm{y}_{k,1:j}\right]
    \label{eq:est_err_cov}
\end{align}
\end{subequations}
and then approximate the belief $\bm{b}(\bm{x}_{k,j})$ by a Gaussian density with mean $\hat{\bm{x}}_{k,j}$ and covariance $\tilde{P}_{k,j}$.
Under this approximation, the estimation error $\tilde{\bm{x}}_{k,j} := \bm{x}_{k,j}-\hat{\bm{x}}_{k,j}$ is conditionally Gaussian with zero mean and covariance $\tilde{P}_{k,j}$.

\begin{figure}[tb]
    \centering
    \resizebox{70mm}{!}{
    \begin{tikzpicture}[
    >=Latex,
    line cap=round,
    line join=round,
    every node/.style={font=\Large},
    state/.style={circle, inner sep=1.7pt, fill},
    traj/.style={very thick, black!70, -{Latex[length=2.3mm]}},
    cov1/.style={blue!65!black, dashed, very thick},
    cov2/.style={red!75!black, dash dot, very thick}
    ]

    \coordinate (xbar)  at (0.00, 0.00);
    \coordinate (xhat)  at (0.50, 1.7);
    \coordinate (xtrue) at (1.3, 1.0);

    \draw[traj] plot[smooth] coordinates {
    (-2.80,-0.9) (xbar) (2.1,0.4) (4.20,0.50)
    };

    \node[black!70, anchor=west] at (2.45,0.8) {Nominal trajectory};

    \draw[cov1, rotate around={80:(xbar)}]
    (xbar) ellipse [x radius=3.7cm, y radius=2.30cm];

    \draw[cov2, rotate around={76:(xhat)}]
    (xhat) ellipse [x radius=1.15cm, y radius=1.45cm];

    \node[state, fill=black!75]        at (xbar)  {};
    \node[state, fill=blue!65!black]   at (xhat)  {};
    \node[state, fill=red!75!black]    at (xtrue) {};

    \node[anchor=south east] at ($(xbar)+(0.1,-0.1)$) {$\bar{\bm{x}}_{k,j}$};
    \node[anchor=south east] at ($(xhat)+(0.1,-0.1)$) {$\hat{\bm{x}}_{k,j}$};
    \node[anchor=south east] at ($(xtrue)+(0.1,-0.5)$) {$\bm{x}_{k,j}$};
    \node[anchor=south east] at ($(xtrue)+(0.45,+0.2)$) {$\tilde{\bm{x}}_{k,j}$};

    \node[inner sep=1pt, text=blue!65!black] at (0.5,4.0) {$\hat{P}_{k,j}$};
    \node[inner sep=1pt, text=red!75!black]  at (0.5,3.16)  {$\tilde{P}_{k,j}$};

    \draw[->, thin, dotted] (xhat) -- (xtrue);

    \end{tikzpicture}
    }
    \caption{Schematic illustration of the nominal state, state estimate, true state, and associated covariances}
    \label{fig:covdef}
\end{figure}
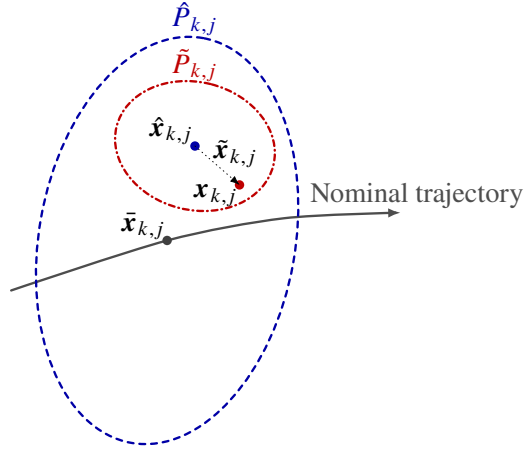

In the mission design phase, future observations have not yet been realized. Therefore, $\hat{\bm{x}}_{k,j}$ is treated as a random variable induced by the process and observation noises, whereas $\tilde{P}_{k,j}$ may also depend on the realized trajectory and observations in the general nonlinear filtering problem.
We further define the nominal state and the state-estimate covariance as
\begin{subequations}\label{eq:planning_moments}
\begin{align}
    \bar{\bm{x}}_{k,j} &:= \mathbb{E}\!\left[\hat{\bm{x}}_{k,j}\right]
    \label{eq:planning_mean_a}\\
    \hat{P}_{k,j} &:= \mathbb{E}\!\left[(\hat{\bm{x}}_{k,j} - \bar{\bm{x}}_{k,j})(\hat{\bm{x}}_{k,j} - \bar{\bm{x}}_{k,j})^\top\right]
    \label{eq:planning_mean_def}
\end{align}
\end{subequations}
and assume that $\hat{\bm{x}}_{k,j}$ also follows a Gaussian distribution, $\hat{\bm{x}}_{k,j} \sim \mathcal{N}\!\left(\bar{\bm{x}}_{k,j},\, \hat{P}_{k,j}\right)$.
By the law of total expectation, the expectation of the true state coincides with that of the state estimate:
\begin{align} \label{eq:mean_property}
    \mathbb{E}\!\left[\bm{x}_{k,j}\right]
    = \mathbb{E}\!\left[\mathbb{E}\!\left[\bm{x}_{k,j} \mid \bar{\bm{u}}_{0:k}, \bm{Y}_{k-1}, \bm{y}_{k,1:j}\right]\right]
    = \mathbb{E}\!\left[\hat{\bm{x}}_{k,j}\right]
    = \bar{\bm{x}}_{k,j}
\end{align}
and the covariance of the true state is given by
\begin{align} \label{eq:true_cov_def}
    \begin{aligned}
    P_{k,j}
    &:= \mathbb{E}\!\left[(\bm{x}_{k,j} - \bar{\bm{x}}_{k,j})(\bm{x}_{k,j} - \bar{\bm{x}}_{k,j})^\top\right] \\
    &= \mathrm{Cov}\!\left[\mathbb{E}\!\left[\bm{x}_{k,j} \mid \bar{\bm{u}}_{0:k}, \bm{Y}_{k-1}, \bm{y}_{k,1:j}\right]\right]
    + \mathbb{E}\!\left[\mathrm{Cov}\!\left[\bm{x}_{k,j} \mid \bar{\bm{u}}_{0:k}, \bm{Y}_{k-1}, \bm{y}_{k,1:j}\right]\right] \\
    &= \hat{P}_{k,j} + \mathbb{E}\!\left[\tilde{P}_{k,j}\right] \\
    &\approx \hat{P}_{k,j} + \tilde{P}^{\mathrm{nom}}_{k,j}
    \end{aligned}
\end{align}
where $\tilde{P}^{\mathrm{nom}}_{k,j}$ is approximated as a deterministic function of the nominal state $\bar{\bm{x}}_{k,j}$, the nominal control $\bar{\bm{u}}_k$, and the assumed process- and observation-noise statistics.
\Eqref{eq:mean_property} and \Eqref{eq:true_cov_def} are consistent with the derivation in Ref.~\cite{Ridderhof2020-cm}.
Under the assumption that the true state lies in the vicinity of the nominal state, we approximate $\tilde{P}_{k,j} \approx \tilde{P}^{\mathrm{nom}}_{k,j}$; that is, the estimation-error covariance obtained from the Bayesian filter is close to its nominal value.
This approximation is justified when the deviation from the nominal trajectory remains sufficiently small.
Therefore, the estimation-error covariance is evaluated along the nominal trajectory and treated deterministically under this approximation.
Figure~\ref{fig:covdef} illustrates the relationships among the true state $\bm{x}_{k,j}$, nominal state $\bar{\bm{x}}_{k,j}$, state estimate $\hat{\bm{x}}_{k,j}$, estimation error $\tilde{\bm{x}}_{k,j}$, estimation-error covariance $\tilde{P}_{k,j}$, and state-estimate covariance $\hat{P}_{k,j}$.

To obtain a deterministic propagation of the first- and second-order moments that represent the approximate belief-state transition model, the prior distributions of the initial state estimate $\hat{\bm{x}}_0$ and the initial estimation error $\tilde{\bm{x}}_0$ are also assumed to be Gaussian, i.e., $\hat{\bm{x}}_0 \sim \mathcal{N}\!\left(\bar{\bm{x}}_0,\, \hat{P}_0\right)$ and $\tilde{\bm{x}}_0 \sim \mathcal{N}\!\left(\bm{0}_{n_x},\, \tilde{P}_0\right)$, where $\bar{\bm{x}}_0$ is the initial nominal state, and $\hat{P}_0$ and $\tilde{P}_0$ denote the initial covariance matrices of the state estimate and the estimation error, respectively.
The quantities $\bar{\bm{x}}_0$, $\hat{P}_0$, and $\tilde{P}_0$ are assumed to be fixed and known.

\subsection{Linearization of dynamics} \label{sec:lin_dyn}
Given a nominal trajectory $(\bar{\bm{x}}(t), \bar{\bm{u}}(t))$, the linearized dynamics describing the evolution of the state deviation $\delta \bm{x}(t)$ are given by
\begin{align}\label{eq:lin_dyn}
    \mathrm{d}\delta \bm{x}(t)
    = \left(\bar{A}(t)\delta \bm{x}(t) + \bar{B}(t)\delta \bm{u}(t)\right)\mathrm{d}t + \bar{G}_x(t)\mathrm{d}\bm{w}_x
\end{align}
where $\bar{A}(t) = \left.\partial \bm{f}/\partial \bm{x}\right|_{\bar{\bm{x}}, \bar{\bm{u}}}$, $\bar{B}(t) = \left.\partial \bm{f}/\partial \bm{u}\right|_{\bar{\bm{x}}, \bar{\bm{u}}}$, and $\bar{G}_x(t) = G_x(\bar{\bm{x}}(t), \bar{\bm{u}}(t))$.
Here, the bilinear terms $\left(\partial G_x/\partial \bm{x} \cdot \delta \bm{x}\right)\mathrm{d}\bm{w}_x$ and $\left(\partial G_x/\partial \bm{u} \cdot \delta \bm{u}\right)\mathrm{d}\bm{w}_x$ are neglected under a small-noise or locally small-deviation assumption.

The nominal control is parameterized as a zero-order-hold continuous thrust input:
\begin{align}
    \bar{\bm{u}}(t) = \bar{\bm{u}}_k, \qquad t \in [t_k, t_{k+1})
\end{align}
Integrating \Eqref{eq:lin_dyn} from $t_{k,j}$ to the next epoch $t_{k,j+1}$ yields the following discrete-time linearized dynamics:
\begin{align}\label{eq:approx_dyn1}
    \delta \bm{x}_{k,j+1} = \bar{A}_{k,j}\delta \bm{x}_{k,j} + \bar{B}_{k,j}\delta \bm{u}_{k,j} + \bar{G}_{x_{k,j}}\bm{w}_{x}
    \qquad j = 0, \ldots, N_j(k)
\end{align}
where $\bm{w}_{x}$ denotes an independent Gaussian random vector with zero mean and identity covariance.
The matrices $\bar{A}_{k,j}$ and $\bar{B}_{k,j}$ are given by
\begin{subequations}\label{eq:lin_dyn_matrices}
\begin{align}
    \bar{A}_{k,j} &:= \frac{\partial \bm{x}_{k,j+1}}{\partial \bm{x}_{k,j}} = \Phi_A(t_{k,j+1}, t_{k,j})
    \label{eq:lin_dyn_matrices_a}\\
    \bar{B}_{k,j} &:= \frac{\partial \bm{x}_{k,j+1}}{\partial \bm{u}_{k,j}}
    = \int_{t_{k,j}}^{t_{k,j+1}} \Phi_A(t_{k,j+1}, t)\bar{B}(t)\mathrm{d}t
    \label{eq:lin_dyn_matrices_b}
\end{align}
\end{subequations}
where $\Phi_A(t_2, t_1)$ denotes the state transition matrix along the nominal trajectory from $t_1$ to $t_2$.

The matrix $\bar{G}_{x_{k,j}}$ is chosen such that the covariance of the discrete process noise $\bar{G}_{x_{k,j}}\bm{w}_{x}$ matches
\begin{align} \label{eq:dyn_noise}
    \bar{Q}_{k,j}
    = \int_{t_{k,j}}^{t_{k,j+1}}
    \Phi_A(t_{k,j+1}, t)\bar{G}_x(t)\bar{G}_x(t)^\top\Phi_A(t_{k,j+1}, t)^\top \mathrm{d}t
\end{align}
where $\bar{G}_x(t)$ is the square-root process-noise matrix evaluated at $(\bar{\bm{x}}(t), \bar{\bm{u}}(t))$, and $\bar{Q}_{k,j} := \bar{G}_{x_{k,j}}\bar{G}_{x_{k,j}}^\top$.
The details of how to compute the matrices $\bar{A}_{k,j}$, $\bar{B}_{k,j}$ are described in the Appendix, and the computation of $\bar{G}_{x_{k,j}}$ is described in \Secref{sec:process_noise}.

\subsection{Linearization of observations and statistical residuals}
Since $\bm{h}_{k,j}$ is assumed to be known, the predicted observation is obtained by evaluating the observation model at the prior state estimate:
\begin{align}\label{eq:obs_eq_hat}
    \hat{\bm{y}}_{{k,j}^-} = \bm{h}_{k,j}(\hat{\bm{x}}_{{k,j}^-}), \qquad j = 1, \ldots, N_j(k)
\end{align}
where $\hat{\bm{x}}_{{k,j}^-}$ is the prior state estimate at time $t_{k,j}$.

By substituting \Eqref{eq:approx_dyn1} and \Eqref{eq:obs_eq_hat} into \Eqref{eq:obs_eq} and linearizing the observation model about the nominal trajectory, the first-order approximation of the observation residual is obtained as
\begin{align}\label{eq:residuals_curr}
    \begin{aligned}
    \delta \bm{y}_{{k,j}^-}
    &:= \bm{y}_{k,j} - \hat{\bm{y}}_{{k,j}^-} \\
    &\approx \bar{C}_{k,j}\tilde{\bm{x}}_{{k,j}^-} + \bar{G}_{y_{k,j}}\bm{w}_{y_{k,j}} \\
    &= \bar{D}_{k,j}\bm{\xi}_{{k,j}^-}
    \end{aligned}
\end{align}
where $\tilde{\bm{x}}_{{k,j}^-} := \bm{x}_{k,j} - \hat{\bm{x}}_{{k,j}^-}$ is the prior estimation error, $\bar{C}_{k,j} = \left.\partial \bm{h}_{k,j}/\partial \bm{x}\right|_{\bar{\bm{x}}_{k,j}}$ is the observation sensitivity matrix, $\bar{G}_{y_{k,j}} = {G}_{y_{k,j}}(\bar{\bm{x}}_{k,j})$, $\bar{D}_{k,j} := [\bar{C}_{k,j} \ \  \bar{G}_{y_{k,j}}]$, and
\begin{align}
    \bm{\xi}_{{k,j}^-}
    :=
    \begin{bmatrix}
        \tilde{\bm{x}}_{{k,j}^-}^\top &
        \bm{w}_{y_{k,j}}^\top
    \end{bmatrix}^\top
\end{align}
collects random quantities affecting the observation residual.
The first- and second-order moments of $\bm{\xi}_{{k,j}^-}$ are given by
\begin{subequations}\label{eq:xi_moments}
\begin{align}
    \mathbb{E}\!\left[\bm{\xi}_{{k,j}^-}\right]
    &= \bm{0}_{n_x + n_{y_{k,j}}}
    \label{eq:xi_moments_a}\\
    P_{\xi_{{k,j}^-}}
    := \mathbb{E}\!\left[\bm{\xi}_{{k,j}^-}\bm{\xi}_{{k,j}^-}^\top\right]
    &=
    \begin{bmatrix}
        \tilde{P}_{{k,j}^-} & O_{n_x \times n_{y_{k,j}}} \\
        O_{n_{y_{k,j}} \times n_x} & I_{n_{y_{k,j}}}
    \end{bmatrix}
    \label{eq:xi_moments_b}
\end{align}
\end{subequations}
and the covariance of the observation residual is defined as
\begin{align}
    P_{{y}_{{k,j}^-}}
    := \mathbb{E}\!\left[\delta \bm{y}_{{k,j}^-} \delta \bm{y}_{{k,j}^-}^{\top}\right]
    = \bar{D}_{k,j} P_{\xi_{{k,j}^-}} \bar{D}_{k,j}^\top
\end{align}

The residual $\delta \bm{y}_{{k,j}^-}$, commonly referred to as the observed-minus-computed (O-C) term in spacecraft orbit determination, is evaluated from the actual observation data and the prior state estimate.
In this study, $\delta \bm{y}_{{k,j}^-}$ is treated as a random variable characterized by $\bm{\xi}_{{k,j}^-}$.

\subsection{Filtered and nominal dynamics} \label{sec:filtered_and_nominal_dyn}

Following Ref.~\cite{Spinello2010-kl}, we adopt the extended Kalman filter (EKF) formulation for state-dependent observation noise derived from Bayes' theorem (cf. \Eqref{eq:Bayesian_filtering}).
The posterior estimate $\hat{\bm{x}}_{k,j+1}$ is given by the maximum a posteriori (MAP) estimate, i.e., the solution of
\begin{subequations}\label{eq:least_square}
\begin{align}
    \hat{\bm{x}}_{k,j+1} &= \argmin_{\bm{\chi}_{k,j+1}} \, l(\bm{\chi}_{k,j+1})
    \label{eq:least_square_a}\\
    l(\bm{\chi}_{k,j+1}) &:= \frac{1}{2}\bm{d}_{k,j+1}^\top \tilde{P}_{{k,j+1}^-}^{-1}\bm{d}_{k,j+1}
    + \frac{1}{2}\bm{e}_{k,j+1}^\top W_{k,j+1}(\bm{\chi}_{k,j+1})\bm{e}_{k,j+1}
    + \frac{1}{2}\log \det W^{-1}_{k,j+1}(\bm{\chi}_{k,j+1}) + \mathrm{const.}
    \label{eq:least_square_b}
\end{align}
\end{subequations}
where $\tilde{P}_{{k,j+1}^-}$ is the prior estimation-error covariance at $t_{k,j+1}$, $\bm{d}_{k,j+1} = \bm{\chi}_{k,j+1} - \hat{\bm{x}}_{{k,j+1}^-}$, and $\bm{e}_{k,j+1} = \bm{y}_{k,j+1} - \bm{h}_{k,j+1}(\bm{\chi}_{k,j+1})$.
The $\log \det W^{-1}_{k,j+1}(\bm{\chi}_{k,j+1})$ term arises from the state dependence of the observation-noise covariance and is absent in the standard state-estimation problem.
$\bm{\chi}_{k,j+1}$ denotes a candidate state in the minimization problem and is introduced to distinguish the optimization variable from the true state $\bm{x}_{k,j+1}$.

Since the observation $\bm{y}_{k,j+1}$ is uncertain and treated as a random variable in the mission design phase, we approximate the state-dependent noise covariance by fixing $W^{-1}_{k,j+1}$ at the nominal state $\bar{\bm{x}}_{k,j+1}$.
We define the fixed observation-noise covariance as
\begin{align}
    \bar{W}^{-1}_{k,j+1}
    := W^{-1}_{k,j+1}(\bar{\bm{x}}_{k,j+1})
    = \bar{G}_{y_{k,j+1}}\bar{G}_{y_{k,j+1}}^\top
\end{align}
Consequently, the MAP estimate is approximated as
\begin{align}\label{eq:state_innovation}
    \hat{\bm{x}}_{k,j+1} = \hat{\bm{x}}_{{k,j+1}^-} + \Lambda_{k,j+1}^{-1}\bm{s}_{k,j+1}
\end{align}
Here, $\Lambda_{k,j+1}$ is the posterior information matrix defined as
\begin{align}\label{eq:fisher_information}
    \Lambda_{k,j+1}
    := \mathbb{E}\!\left[\frac{\partial^2 l(\bm{\chi}_{k,j+1})}{\partial \bm{\chi}_{k,j+1}\partial \bm{\chi}_{k,j+1}^\top}\right]
    = \tilde{P}_{{k,j+1}^-}^{-1} + \bar{S}_{k,j+1}
\end{align}
and $\bm{s}_{k,j+1}$ and $\bar{S}_{k,j+1}$ are given by
\begin{subequations}\label{eq:score_terms}
\begin{align}
    \bm{s}_{k,j+1}
    &= \bar{C}_{k,j+1}^\top \bar{W}_{k,j+1}\bm{e}_{k,j+1}
    \label{eq:score}\\
    \bar{S}_{k,j+1}
    &= \bar{C}_{k,j+1}^\top \bar{W}_{k,j+1}\bar{C}_{k,j+1}
    \label{eq:score2}
\end{align}
\end{subequations}

The prior state estimate and estimation-error covariance at $t_{k,j+1}$ are propagated from the posterior quantities at $t_{k,j}$ as
\begin{subequations}\label{eq:prior_propagation}
\begin{align}
    \hat{\bm{x}}_{{k,j+1}^-}
    &= \bm{f}_{k,j}(\hat{\bm{x}}_{k,j}, \bm{u}_{k,j})
    \label{eq:prior_state_esimate}\\
    \tilde{P}_{{k,j+1}^-}
    &= \bar{A}_{k,j}\tilde{P}_{k,j}\bar{A}_{k,j}^\top + \bar{G}_{x_{k,j}}\bar{G}_{x_{k,j}}^\top
    \label{eq:prior_est_err_cov}
\end{align}
\end{subequations}
where $\bm{f}_{k,j}(\hat{\bm{x}}_{k,j}, \bm{u}_{k,j})$ denotes the discrete-time nonlinear dynamics obtained by numerically integrating the deterministic part of \Eqref{eq:stochastic_dynamics}.
Under the Gaussian-belief assumption, the MAP estimate in \Eqref{eq:state_innovation} coincides with the conditional expectation $\mathbb{E}\!\left[\bm{x}_{k,j+1} \mid \bar{\bm{u}}_{0:k}, \bm{Y}_{k-1}, \bm{y}_{k,1:j+1}\right]$, because the posterior is Gaussian and its mean and mode are identical. Note that this equivalence does not hold in general for non-Gaussian posteriors.
Since all terms in \Eqref{eq:state_innovation}, \Eqref{eq:fisher_information}, and \Eqref{eq:score_terms} are evaluated along the nominal trajectory, e.g., at $\bar{\bm{x}}_{k,j}$ and $\bar{\bm{x}}_{k,j+1}$, the resulting propagation is more accurately described as a linearized Kalman filter than as the standard EKF, in which the matrices are evaluated at the latest state estimate $\hat{\bm{x}}_{k,j}$.

Under the first-order approximation $\bm{e}_{k,j+1} \approx \delta \bm{y}_{k,j+1}^-$, substituting \Eqref{eq:residuals_curr} into \Eqref{eq:score} yields
\begin{align}\label{eq:map_gradient}
    \bm{s}_{k,j+1}
    = \bar{C}_{k,j+1}^\top \bar{W}_{k,j+1}\bar{D}_{k,j+1}\bm{\xi}_{{k,j+1}^-}
\end{align}
Accordingly, the intermediate propagation with state estimation can be written as
\begin{align}\label{eq:filtered_dynamics}
    \hat{\bm{x}}_{k,j+1}
    = \bm{f}_{k,j}(\hat{\bm{x}}_{k,j}, \bm{u}_{k,j})
    + \tilde{P}_{k,j+1}\bar{C}_{k,j+1}^\top \bar{W}_{k,j+1}\bar{D}_{k,j+1}\bm{\xi}_{{k,j+1}^-}
\end{align}
and the propagation of the estimation-error covariance is then approximated by
\begin{align}\label{eq:propagation_est_err_cov}
    \begin{aligned}
    \tilde{P}_{k,j+1}
    &\approx \Lambda_{k,j+1}^{-1}
    &= \left[\left(\bar{A}_{k,j}\tilde{P}_{k,j}\bar{A}_{k,j}^\top + \bar{G}_{x_{k,j}}\bar{G}_{x_{k,j}}^\top\right)^{-1} + \bar{S}_{k,j+1}\right]^{-1}
    \end{aligned}
\end{align}

By taking the expectation of both sides of \Eqref{eq:filtered_dynamics} and applying the first-order approximation, the nominal dynamics are approximated as
\begin{align}\label{eq:nominal_dynamics}
    \begin{aligned}
        \bar{\bm{x}}_{k,j+1}
        &=
        \mathbb{E}\!\left[
            \bm{f}_{k,j}(\hat{\bm{x}}_{k,j}, \  \bm{u}_{k,j})
        \right]
        +
        \tilde{P}_{k,j+1}
        \bar{C}_{k,j+1}^{\top}
        \bar{W}_{k,j+1}
        \bar{D}_{k,j+1}
        \mathbb{E}\!\left[
            \bm{\xi}_{k,j+1}^-
        \right] \\
        &\approx
        \bm{f}_{k,j}
        \left(
            \mathbb{E}\!\left[\hat{\bm{x}}_{k,j}\right],
            \mathbb{E}\!\left[\bm{u}_{k,j}\right]
        \right) \\
        &=
        \bm{f}_{k,j}(\bar{\bm{x}}_{k,j}, \  \bar{\bm{u}}_k)
    \end{aligned}
\end{align}
where $\bar{\bm{x}}_{k,j+1} := \mathbb{E}\!\left[\hat{\bm{x}}_{k,j+1}\right]$ and $\mathbb{E}\!\left[\bm{u}_{k,j}\right]=\bar{\bm{u}}_k$ because $\mathbb{E}\!\left[\delta\hat{\bm{x}}_{k,j}\right]=\bm{0}_{n_x}$.

\subsection{Belief-state transition model} \label{sec:planning_belief_dyn}
Although the nominal control $\bar{\bm{u}}_k$ and feedback gain $K_k$ are held fixed within each stage, the feedback correction is recomputed at each intermediate epoch using the latest state estimate.
Thus, for $t \in [t_{k,j}, t_{k,j+1})$, the applied control is modeled as
\begin{align}
    \begin{aligned} \label{eq:controller}
    \bm{u}_{k,j} &= \bar{\bm{u}}_k + K_k(\hat{\bm{x}}_{k,j}-\bar{\bm{x}}_{k,j}) \\
     &= \bar{\bm{u}}_k + K_k \delta \hat{\bm{x}}_{k,j}
    \end{aligned}
\end{align}
Because $\mathbb{E}\!\left[\delta\hat{\bm{x}}_{k,j}\right]=\bm{0}_{n_x}$, the expected applied control satisfies $\mathbb{E}\!\left[\bm{u}_{k,j}\right]=\bar{\bm{u}}_k$.
The control deviation is defined as $\delta \bm{u}_{k,j} := \bm{u}_{k,j} - \bar{\bm{u}}_k$ and its covariance is given by
\begin{align}
    P_{u_{k,j}} := \mathbb{E}\!\left[\delta \bm{u}_{k,j}  \delta \bm{u}_{k,j}^\top \right] = K_k \hat{P}_{k,j} K_k^\top
\end{align}

By substituting \Eqref{eq:controller} into \Eqref{eq:filtered_dynamics} and expanding to first order about the nominal trajectory, the propagation of the estimate deviation is approximated as
\begin{align}
    \begin{aligned}
    \delta \hat{\bm{x}}_{k,j+1}
    &\approx \bar{A}_{k,j}\delta \hat{\bm{x}}_{k,j} + \bar{B}_{k,j}\delta \bm{u}_{k,j} + \tilde{P}_{k,j+1}\bm{s}_{k,j+1} \\
    &= \left(\bar{A}_{k,j} + \bar{B}_{k,j}K_k\right)\delta \hat{\bm{x}}_{k,j} + \tilde{P}_{k,j+1}\bm{s}_{k,j+1}
    \end{aligned}
\end{align}
The propagation of the state-estimate covariance is then given by
\begin{align} \label{eq:planning_cov_dyn}
    \hat{P}_{k,j+1}
    = \mathcal{A}_{k,j}\hat{P}_{k,j}\mathcal{A}_{k,j}^\top
    + \mathcal{F}_{k,j+1}P_{\xi_{{k,j+1}^-}}\mathcal{F}_{k,j+1}^\top
\end{align}
where
\begin{subequations}\label{eq:planning_cov_terms}
\begin{align}
    \mathcal{A}_{k,j} &= \bar{A}_{k,j} + \bar{B}_{k,j}K_k
    \label{eq:planning_cov_terms_a}\\
    \mathcal{F}_{k,j+1} &= \tilde{P}_{k,j+1} \bar{C}_{k,j+1}^\top \bar{W}_{k,j+1} \bar{D}_{k,j+1}
    \label{eq:planning_cov_terms_b}
\end{align}
\end{subequations}

By vectorizing the belief state, we define the augmented state and control as
\begin{subequations}
\begin{align}
    \bm{X}_{k,j}
    &=
    \begin{bmatrix}
        \bar{\bm{x}}_{k,j}^\top &
        \mathrm{vec}(\tilde{P}_{k,j})^\top &
        \mathrm{vec}(\hat{P}_{k,j})^\top
    \end{bmatrix}^\top \\
    \bm{U}_k &=
    \begin{bmatrix}
        \bar{\bm{u}}_k^\top &
        \mathrm{vec}(K_k)^\top
    \end{bmatrix}^\top
\end{align}
\end{subequations}
where the feedback gain $K_k$ is included in the augmented control to explicitly consider $\hat{P}_k$ as controlled variables for the future distribution.
Accordingly, the intermediate belief-state transition model can be written as
\begin{align} \label{eq:belief_space_dyn}
    \bm{X}_{k,j+1} &= \bm{F}_{k,j}\!\left(\bm{X}_{k,j}, \bm{U}_k\right), & j = 0, \ldots, N_j(k)
\end{align}
where $\bm{F}_{k,j}\!\left(\bm{X}_{k,j}, \bm{U}_k\right)$ concatenates the transition model of the nominal state, the estimation-error covariance, and the state-estimate covariance, described in \Eqsref{eq:nominal_dynamics}, (\ref{eq:propagation_est_err_cov}), and (\ref{eq:planning_cov_dyn}), respectively.
If no observation is associated with an epoch $t_{k,j+1}$, including the case $N_j(k)=0$, the corresponding measurement-update terms are omitted by setting $\bar{S}_{k,j+1}=O_{n_x}$ and removing the term $\mathcal{F}_{k,j+1}P_{\xi_{{k,j+1}^-}}\mathcal{F}_{k,j+1}^\top$ from the propagations of $\tilde{P}_{k,j+1}$ and $\hat{P}_{k,j+1}$, respectively.

\begin{figure}[bt]
    \centering
    \begin{tikzpicture}[
        font=\normalsize,
        >={Latex[length=1.8mm,width=1.3mm]},
        edge/.style={-Latex, line width=0.55pt},
        edgeLab/.style={midway, fill=white, inner sep=0.6pt, font=\scriptsize},
        state/.style={draw, minimum width=8.5mm, minimum height=6.5mm, inner sep=1pt, align=center},
        action/.style={draw, minimum width=8.5mm, minimum height=6.5mm, inner sep=1pt, align=center},
        obs/.style={draw, minimum width=8.5mm, minimum height=6.5mm, inner sep=1pt, align=center},
        latent/.style={state, fill=black!3}
    ]

    \node[latent] (xk) {$\bm{X}_k$};

    \node[latent, right=16mm of xk] (xkj) {$\bm{X}_{k,j}$};
    \node[latent, right=12mm of xkj] (xkj1) {$\bm{X}_{k,j+1}$};

    \node[font=\footnotesize, left=4mm of xkj] (dotsL) {$\cdots$};
    \node[font=\footnotesize, right=4mm of xkj1] (dotsR) {$\cdots$};

    \node[latent, right=16mm of xkj1] (xkp) {$\bm{X}_{k+1}$};

    \draw[edge] (xkj) -- (xkj1)
    node[edgeLab, above] {$\bm{F}_{k,j}$};

    \draw[edge] (dotsL) -- (xkj);
    \draw[edge] (xkj1) -- (dotsR);

    \begin{scope}[on background layer]
    \node[
        draw,
        rounded corners=0pt,
        inner sep=2pt,
        fit=(dotsL)(xkj)(xkj1)(dotsR),
        label={[font=\scriptsize]north:$\bm{F}_k(\bm{X}_k,\bm{U}_k)$}
    ] (dyn) {};
    \end{scope}

    \node[action, above=7mm of xk] (uk) {$\bm{U}_k$};

    \draw[edge] (xk.east) -- (dyn.west);
    \draw[edge] (uk.east) -| ($(dyn.north)+(-16mm,0)$);
    \draw[edge] (dyn.east) -- (xkp.west);

    \end{tikzpicture}
    \caption{Block diagram of the sequential propagation of the belief-state dynamics}
    \label{fig:propagation_block}
\end{figure}
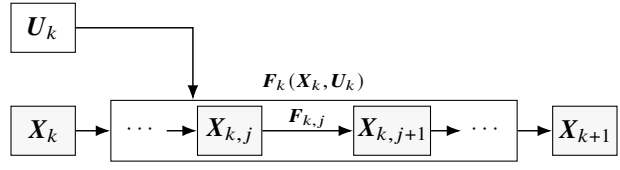

For use in trajectory optimization algorithms, the stage-to-stage belief-state transition must be written directly as a function of $\bm{X}_k$ and $\bm{U}_k$, because $\bm{X}_{k,j}$ is an intermediate variable determined by these quantities.
Letting $\bm{X}_k := \bm{X}_{k,0}$, the belief-state transition model from $t_k$ to $t_{k+1}$ is obtained by sequentially applying \Eqref{eq:belief_space_dyn} over the intermediate epochs:
\begin{align} \label{eq:dyn_aug}
    \bm{X}_{k+1} = \bm{F}_k\!\left(\bm{X}_k, \bm{U}_k\right)
\end{align}
Figure~\ref{fig:propagation_block} illustrates the sequential state-propagation process described by \Eqref{eq:dyn_aug}.

To obtain the posterior distribution $p(\bm{x}_{k+1} \mid \bm{Y}_k, \bm{U}_k)$, Bayes' rule may also be applied directly, as in Ref.~\cite{Indelman2015-jb}.
In that case, only the distribution linking $\bm{x}_k$ and $\bm{x}_{k+1}$ conditioned on the observation sequence $\bm{Y}_k$ is of interest, since all intermediate states $\bm{x}_{k,j}$ can be marginalized out.
However, when the dynamical system contains process noise, filtering conditioned on $\bm{Y}_k$ induces strong correlations among observations, as shown in Ref.~\cite[Eq.~(12)]{Carpenter2023-qo}.
These correlations lead to cumbersome algorithmic implementation, particularly in the partial derivatives of the belief-state transition model.
For this reason, sequential propagation with intermediate states is adopted in this study.

The dimensions of the augmented state and control are $n_X = n_x + 2n_x^2$ and $n_U = n_u + n_u n_x$, respectively.
Using square-root factors of the covariance matrices exploits symmetric structure and reduces the dimension of the augmented state to $n_x + n_x(n_x+1)$.
Although square-root representations may improve numerical stability and reduce dimensionality, we deliberately retain the full covariance matrices in this study because evaluating the state transition matrices for the belief-state transition model is simpler than with square-root factors, which require QR decompositions and second-order differentiation through them.

\section{Partially Observable Stochastic Differential Dynamic Programming} \label{sec:po_sddp}
This section presents the PO-SDDP algorithm for solving partially observable trajectory optimization problems.
The algorithm applies a DDP-based method with an augmented Lagrangian formulation to the deterministic belief-state transition model derived in the previous section.

For the nonlinear partially observable problem considered in this study, the separation principle does not generally decouple state estimation from trajectory optimization.
The covariance dynamics are evaluated along the nominal trajectory being optimized; therefore, their coefficients vary with the nominal state and control inputs.
This coupling appears through matrices such as $\bar{A}_{k,j}$, $\bar{B}_{k,j}$, $\bar{G}_{x_{k,j}}$, $\bar{C}_{k,j+1}$, $\bar{S}_{k,j+1}$, $\mathcal{A}_{k,j}$, and $\mathcal{F}_{k,j+1}$, which depend on $(\bar{\bm{x}}_{k,j}, \bar{\bm{u}}_k)$.
Consequently, the augmented dynamics in \Eqref{eq:belief_space_dyn} must be differentiated with respect to the full augmented state and control.
The DDP backward pass then uses the first- and second-order derivatives of this augmented transition map to construct local quadratic subproblems.

\subsection{Problem definition}
Consider the following optimization problem:
\begin{subequations}
\begin{align} \label{eq:org_prb_bgn}
    \min_{\bm{U}_0, \bm{U}_1, \ldots, \bm{U}_{N-1}} \quad
    & \sum^{N-1}_{k=0} L_k(\bm{X}_k, \bm{U}_k) + \varphi(\bm{X}_N) \\
    \mathrm{s.t.} \quad
    & \bm{X}_{k+1} = \bm{F}_k(\bm{X}_k, \bm{U}_k) \label{eq:org_prb_dyn}\\
    & \bm{c}_{k,\mathcal{I}}(\bm{X}_k, \bm{U}_k) \leq \bm{0}_{n_{{c}_{k,\mathcal{I}}}} \\
    & \bm{c}_{k,\mathcal{E}}(\bm{X}_k, \bm{U}_k) = \bm{0}_{n_{{c}_{k,\mathcal{E}}}} \\
    & \bm{\phi}(\bm{X}_N) \leq \bm{0}_{n_\phi} \\
    & \bm{\psi}(\bm{X}_N) = \bm{0}_{n_\psi} \label{eq:org_prb_end}
\end{align}
\end{subequations}
where $L_k(\bm{X}_k, \bm{U}_k) : \mathbb{R}^{n_X} \times \mathbb{R}^{n_U} \rightarrow \mathbb{R}$ and $\varphi(\bm{X}_N) : \mathbb{R}^{n_X} \rightarrow \mathbb{R}$ denote the stage and terminal cost functions, respectively, and $\bm{X}_k$ and $\bm{U}_k$ denote the augmented state and augmented control at $t_k$.
The functions $\bm{c}_{k,\mathcal{I}}(\bm{X}_k, \bm{U}_k)  : \mathbb{R}^{n_X} \times \mathbb{R}^{n_U} \rightarrow \mathbb{R}^{n_{{c}_{k,\mathcal{I}}}}$ and $\bm{c}_{k,\mathcal{E}}(\bm{X}_k, \bm{U}_k) : \mathbb{R}^{n_X} \times \mathbb{R}^{n_U} \rightarrow \mathbb{R}^{n_{{c}_{k,\mathcal{E}}}}$ denote the stage inequality and equality constraints, and the functions $\bm{\phi}(\bm{X}_N) : \mathbb{R}^{n_X} \rightarrow \mathbb{R}^{n_\phi}$ and $\bm{\psi}(\bm{X}_N) : \mathbb{R}^{n_X}\rightarrow \mathbb{R}^{n_\psi}$ represent the terminal inequality and equality constraints, respectively.

\subsection{Augmented Lagrangian method}
Although various approaches have been developed to handle constraints in DDP, including active-set methods~\cite{Lantoine2012-ak, Xie2017-ma} and interior-point DDP~\cite{Pavlov2021-ll}, the proposed method adopts the augmented Lagrangian DDP (AL-DDP) framework~\cite{howell2019altro, Pellegrini2020-ep}.

In AL-DDP, the stage and terminal constraints are incorporated into the cost function. The resulting augmented cost function is defined as
\begin{align} \label{eq:aug_cost}
    J_0 = \sum_{k=0}^{N-1} \tilde{L}_k(\bm{X}_k, \bm{U}_k) + \tilde{\varphi}(\bm{X}_N)
\end{align}
with
\begin{subequations}
\begin{align}
    \tilde{L}_k
    &=
    L_k(\bm{X}_k, \bm{U}_k)
    + \mathcal{P}_{\mathcal{I}} \left(\bm{c}_{k,\mathcal{I}}(\bm{X}_k,\bm{U}_k), \bm{\lambda}_{k,\mathcal{I}}, \sigma_{k,\mathcal{I}}\right) + \mathcal{P}_{\mathcal{E}} \left(\bm{c}_{k,\mathcal{E}}(\bm{X}_k,\bm{U}_k), \bm{\lambda}_{k,\mathcal{E}}, \sigma_{k,\mathcal{E}}\right) \\
    \tilde{\varphi}(\bm{X}_N)
    &=
    \varphi(\bm{X}_N)
    + \mathcal{P}_{\mathcal{I}} \left(\bm{\phi}(\bm{X}_N), \bm{\lambda}_{N,\mathcal{I}}, \sigma_{N,\mathcal{I}}\right) + \mathcal{P}_{\mathcal{E}} \left(\bm{\psi}(\bm{X}_N), \bm{\lambda}_{N,\mathcal{E}}, \sigma_{N,\mathcal{E}}\right)
\end{align}
\end{subequations}
where $\bm{\lambda}_{k,\mathcal{I}} \in \mathbb{R}^{n_{{c}_{k,\mathcal{I}}}}$ and $\bm{\lambda}_{k,\mathcal{E}} \in \mathbb{R}^{n_{{c}_{k,\mathcal{E}}}}$ denote the Lagrange multipliers for the stage constraints, and $\bm{\lambda}_{N,\mathcal{I}} \in \mathbb{R}^{n_{\phi}}$ and $\bm{\lambda}_{N,\mathcal{E}} \in \mathbb{R}^{n_{\psi}}$ denote the Lagrange multipliers for the terminal constraints. The positive scalars $\sigma_{k,\mathcal{I}}$, $\sigma_{k,\mathcal{E}}$, $\sigma_{N,\mathcal{I}}$, and $\sigma_{N,\mathcal{E}}$ denote the corresponding penalty parameters.
The augmented Lagrangian penalty for equality constraints is defined as
\begin{align}
\mathcal{P}_{\mathcal{E}}(\bm{c},\bm{\lambda},\sigma) =
\bm{\lambda}^{\top}\bm{c} + \frac{\sigma}{2} \bm{c}^{\top} \bm{c}
\end{align}
For a generic inequality constraint vector $\bm{c}\in\mathbb{R}^{n_c}$, the penalty for inequality constraints is defined componentwise as
\begin{align}
\mathcal{P}_{\mathcal{I}}(\bm{c},\bm{\lambda},\sigma) = \sum_{i=1}^{n_c} \left(\lambda_i c_i + \frac{1}{2} a_i c_i^2 \right)
\end{align}
with
\begin{align}
a_i =
\begin{cases}
\sigma, & c_i \geq 0 \ \text{or} \ \lambda_i > 0 \\
0, & \text{otherwise}
\end{cases}
\end{align}

The AL-DDP algorithm consists of nested inner and outer loops.
In the inner loop, DDP locally optimizes the control sequence by solving the unconstrained problem defined by the augmented cost function via backward and forward passes.
After the inner loop has approximately converged, the outer loop updates the Lagrange multipliers and penalty parameters.
The Lagrange multipliers and penalty parameters are updated as
\begin{subequations}
\begin{align}
\bm{\lambda}_{k,\mathcal{I}}^{+} &= \max \left(\bm{0}_{n_{{c}_{k,\mathcal{I}}}}, \bm{\lambda}_{k,\mathcal{I}} + \sigma_{k,\mathcal{I}} \bm{c}_{k,\mathcal{I}}\right), &
\bm{\lambda}_{k,\mathcal{E}}^{+} &= \bm{\lambda}_{k,\mathcal{E}} + \sigma_{k,\mathcal{E}} \bm{c}_{k,\mathcal{E}} \\
\sigma^{+}_{k,\mathcal{I}} &= \gamma_{k,\mathcal{I}} \sigma_{k,\mathcal{I}}, & \sigma^{+}_{k,\mathcal{E}} &= \gamma_{k,\mathcal{E}} \sigma_{k,\mathcal{E}}
\end{align}
\end{subequations}
where the maximum operation is applied componentwise, and $\gamma_{k,\mathcal{I}}$ and $\gamma_{k,\mathcal{E}}$ are penalty scaling constants.

\subsection{Backward pass}
Given the augmented cost function in \Eqref{eq:aug_cost}, the value function satisfies the Bellman recursion
\begin{align}
 J^*_k (\bm{X}_{k}) = \min_{\bm{U}_k} \left[\tilde{L}_k (\bm{X}_k, \bm{U}_k) + J^*_{k+1} (\bm{F}_k\!\left(\bm{X}_k, \bm{U}_k\right)) \right]
\end{align}
with the terminal condition $J_N^* (\bm{X}_N)=\tilde{\varphi}(\bm{X}_N)$, where $J^*_{k+1}(\bm{F}_k\!\left(\bm{X}_k, \bm{U}_k\right))$ denotes the optimal cost-to-go, representing the future cost along the trajectory controlled by the optimal policy.
In the backward pass, the control variations are obtained through minimizing the quadratic expansion of the cost-to-go function from the terminal stage to the initial one.

By introducing the state $\bm{Z}_k=[\bm{X}_k^\top \ \  \bm{U}_k^\top]^\top$ that combines $\bm{X}_k$ with $\bm{U}_k$, the quadratic expansion of the combined belief-state transition model can be written as
\begin{align} \label{eq:dyn_expansion}
    \delta \bm{Z}_{k+1} = \Phi^1_k \delta \bm{Z}_k + \frac{1}{2} \delta \bm{Z}_k^\top \cdot \Phi^2_k \delta \bm{Z}_k
\end{align}
where $\Phi^1_k$ and $\Phi^2_k$ denote the first- and second-order state transition matrices (STMs) from $t_k$ to $t_{k+1}$, and $(\cdot)$ denotes a vector-tensor contraction.
The STMs are associated with the combined dynamics defined as
\begin{align}
    \bm{F}_{Z,k}(\bm{X}_k, \bm{U}_k) =
    \begin{bmatrix}
        \bm{F}_k(\bm{X}_k, \bm{U}_k)^\top & \bm{U}_k^\top
    \end{bmatrix}^\top
\end{align}
since the controls are modeled as the zero-order-hold continuous thrust input.
By expanding the cost-to-go function up to second order, the derivatives of the local quadratic expansion are written as
\begin{subequations}
\begin{align}
J_{Z,k} &= \tilde{L}_{Z,k} + \Phi_k^{1\top} J^*_{X,k+1} \\
J_{ZZ,k} &= \tilde{L}_{ZZ,k} + \Phi_k^{1\top} J^*_{XX,k+1} \Phi_k^1 + J_{X,k+1}^{*\top} \cdot \Phi_k^2
\end{align}
\end{subequations}
where
\begin{align}
J_{Z,k} &= \begin{bmatrix}
J_{X,k} \\
J_{U,k}
\end{bmatrix}, &
J_{ZZ,k} =
\begin{bmatrix}
J_{XX,k} & J_{XU,k} \\
J_{UX,k} & J_{UU,k}
\end{bmatrix}
\end{align}

If $J_{UU,k}$ is not positive definite, the resulting control update may fail to satisfy a descent condition.
To address this issue, the algorithm incorporates a trust-region method that regularizes $J_{UU,k}$ and restricts the control update $\delta \bm{U}_k$ to a region where the quadratic approximation remains valid.
The trust-region subproblem is given by
\begin{subequations} \label{eq:trqp}
\begin{align}
    \min_{\delta \bm{U}_k} \quad
    & J_{U,k}^\top \delta \bm{U}_k + \frac{1}{2} \delta \bm{U}_k^\top J_{UU,k} \delta \bm{U}_k \\
    \mathrm{s.t.} \quad
    & \lVert D_{\mathrm{tr}} \delta \bm{U}_k \rVert \leq \Delta
\end{align}
\end{subequations}
where $\lVert\cdot\rVert$ denotes the Euclidean norm, $D_{\mathrm{tr}}$ is a positive definite scaling matrix and $\Delta$ is the trust-region radius.
The matrix $D_{\mathrm{tr}}$ defines a hyperellipsoid in the control space of $\bm{U}_k$.
Using the regularized Hessian $\tilde{J}_{UU,k} = J_{UU,k} + \gamma D_{\mathrm{tr}}^\top D_{\mathrm{tr}}$ where $\gamma$ is the Lagrange multiplier associated with the trust-region constraint, the optimal local control update is obtained as
\begin{align} \label{eq:local_update_ctrl}
    \begin{aligned}
    \delta \bm{U}_k^* &= -\tilde{J}_{UU,k}^{-1} \left( J_{U,k} + J_{UX,k} \delta \bm{X}_k \right) \\
    &= \bm{\alpha}_k + \beta_k \delta \bm{X}_k
    \end{aligned}
\end{align}
where $\bm{\alpha}_k$ is the feedforward controller and $\beta_k$ is the feedback-gain matrix defined by $\bm{\alpha}_k = -\tilde{J}_{UU,k}^{-1} J_{U,k}$ and $\beta_k = -\tilde{J}_{UU,k}^{-1} J_{UX,k}$, respectively.

The expected cost reduction and value-function derivatives passed to the preceding stage are obtained by substituting the local policy into the quadratic expansion:
\begin{subequations}
\begin{align}
\mathrm{ER}_k &= \mathrm{ER}_{k+1} + J_{U,k}^\top \bm{\alpha}_k + \frac{1}{2} \bm{\alpha}_k^\top J_{UU,k} \bm{\alpha}_k \\
J^*_{X,k} &= J_{X,k} + J_{U,k}^\top \beta_k + \bm{\alpha}_k^\top J_{UX,k} + \bm{\alpha}_k^\top J_{UU,k} \beta_k \\
J^*_{XX,k} &= J_{XX,k} + \beta_k^\top J_{UX,k} + J_{UX,k}^\top \beta_k + \beta_k^\top J_{UU,k} \beta_k
\end{align}
\end{subequations}
with terminal conditions $\mathrm{ER}_{N} = 0$, $J_{X,N}^* = \tilde{\varphi}_{X}$, and $J_{XX,N}^* = \tilde{\varphi}_{XX}$.

\subsection{Forward pass}
After the backward pass, a candidate reference trajectory is generated using the latest control policy.
Given the reference state and control at the $i$th iteration, the candidate reference trajectory for the $(i+1)$th iteration is computed by forward belief propagation:
\begin{subequations}\label{eq:forward_pass}
\begin{align}
    {\bm{X}}^{(i+1)}_{k+1} &= \bm{F}_k\left({\bm{X}}^{(i+1)}_k, {\bm{U}}^{(i+1)}_k\right) \\
    {\bm{U}}^{(i+1)}_k  &= {\bm{U}}^{(i)}_k + \bm{\alpha}_k + \beta_k \left({\bm{X}}^{(i+1)}_k - {\bm{X}}^{(i)}_k\right)
\end{align}
\end{subequations}
The total cost of the updated reference trajectory is then evaluated as
\begin{align} \label{eq:new_cost}
 J^{(i+1)}_0 = \sum_{k=0}^{N-1} \tilde{L}_k\left({\bm{X}}^{(i+1)}_k, {\bm{U}}^{(i+1)}_k\right) + \tilde{\varphi}\left({\bm{X}}^{(i+1)}_N\right)
\end{align}
The ratio of the actual cost reduction $J^{(i+1)}_0 - J^{(i)}_0$ to the expected reduction $\mathrm{ER}_0$ is then used to determine whether the candidate reference trajectory is accepted.
If the candidate is accepted, the algorithm proceeds to the next iteration by reevaluating the quadratic expansions of the belief-state transition model $\bm{F}_{Z,k}$, stage cost $\tilde{L}_k$, and terminal cost $\tilde{\varphi}$ around the updated reference trajectory $\left({\bm{X}}^{(i+1)}_k, {\bm{U}}^{(i+1)}_k\right)$.
Otherwise, the backward pass is repeated with a reduced trust-region radius $\Delta$ while reusing previously computed derivatives until the control update is accepted.

\section{Implementation Details} \label{sec:impl_details}
In this section, we describe implementation details that enhance the numerical stability and computational efficiency of the proposed framework.
Specifically, we compute the first- and second-order derivatives of the covariance dynamics semi-analytically by reusing derivatives of the nominal dynamics, and introduce representative cost and constraint formulations for space mission design, smoothed by using Schatten-norm surrogates.

\subsection{Partial derivatives of covariance propagation} \label{sec:imp_cov_derivs}
The most computationally expensive part of the framework is the computation of the first- and second-order STMs required for covariance propagation.
Direct application of finite differences or automatic differentiation (AD) is computationally prohibitive because of the high dimensionality of the augmented dynamics.
By exploiting the dependence of the covariance matrices on the nominal state and control, the STMs for covariance propagation can instead be computed from analytic derivatives expressed in terms of the STMs of the nominal-state dynamics.
For example, the derivatives of the prior estimation-error covariance with respect to the nominal state can be written in terms of tensor multiplications as
\begin{align} \label{eq:prior_est_err_derivative}
    \frac{\partial \tilde{P}^{a b}_{{k,j+1}^-}}{\partial x^c_{k,j}} &=
    \frac{\partial \bar{A}^{a \alpha}_{{k,j}}}{\partial x^c_{k,j}} \tilde{P}^{\alpha \beta}_{k,j} \bar{A}^{b \beta}_{k,j}
    + \bar{A}^{a \alpha}_{k,j} \tilde{P}^{\alpha \beta}_{k,j} \frac{\partial \bar{A}^{b \beta}_{{k,j}}}{\partial x^c_{k,j}}
    + \frac{\partial \bar{G}^{a \alpha}_{x_{k,j}}}{\partial x^c_{k,j}} \bar{G}^{b \alpha}_{x_{k,j}}
    + \bar{G}^{a \alpha}_{x_{k,j}} \frac{\partial \bar{G}^{b \alpha}_{x_{k,j}}}{\partial x^c_{k,j}}
\end{align}
where repeated superscripts follow the summation convention.
On the right-hand side, the derivative of $\bar{A}_{k,j}$ with respect to $x_{k,j}$ corresponds to the second-order STMs of the nominal dynamics, whereas the derivative of $\bar{G}_{x_{k,j}}$ with respect to $x_{k,j}$ can be obtained through AD\@.

The procedure for computing the STMs of the nominal dynamics is summarized in the Appendix.
Because these derivatives correspond to the belief-state transition over each intermediate interval from $t_{k,j}$ to $t_{k,j+1}$, the Appendix also describes how the intermediate STMs are composed to obtain the stage-to-stage STMs.

In the second-order derivatives of covariance propagation, the third-order derivatives of the nominal-state dynamics emerge, for example,
\begin{align}
    \frac{\partial^2 \bar{A}^{a b}_{{k,j}}}{\partial x^c_{k,j} \partial x^d_{k,j}} = \frac{\partial^3 x^{a}_{{k,j+1}}}{\partial x^b_{k,j} \partial x^c_{k,j} \partial x^d_{k,j}}
\end{align}
The current implementation omits these terms to reduce computational burden.
The effect of this approximation is mitigated in practice by the trust-region method in the backward pass, which limits the update when the quadratic expansion does not accurately predict the cost reduction.

Accordingly, the problem-dependent implementation effort is largely reduced to specifying the dynamics, observation models, and uncertainty models.
The corresponding model-dependent derivatives can be evaluated using AD, whereas the tensor operations associated with covariance propagation are handled independently of the model.

\subsection{Approximation of discretized process noise} \label{sec:process_noise}
As described in \Secref{sec:lin_dyn}, the discretized process-noise matrix $\bar{G}_{x_{k,j}}$ must be computed in the discretization procedure.
This section introduces the approximation approach used in this study.

A direct approach is to integrate the continuous-time Lyapunov equation for $\bar{Q}(t)$:
\begin{align} \label{eq:lyapunov}
    \dot{\bar{Q}}(t) = \bar{A}(t)\bar{Q}(t) + \bar{Q}(t)\bar{A}^\top(t) + \bar{G}(t)\bar{G}^\top(t)
\end{align}
from $t_{k,j}$ to $t_{k,j+1}$ with the initial condition $\bar{Q}(t_{k,j}) = O_{n_x}$.
The matrix $\bar{G}_{x_{k,j}}$ can then be obtained from a Cholesky decomposition of $\bar{Q}_{k,j}$.
Because the additional numerical integration of $\bar{Q}(t)$ increases the computational cost, the PO-SDDP algorithm instead uses the trapezoidal rule to approximate $\bar{G}_{x_{k,j}}$.
Using the first-order STM $\bar{A}_{k,j}$, \Eqref{eq:dyn_noise} is approximated as
\begin{align}
    \bar{Q}_{k,j} \approx \frac{\Delta t_{k,j}}{2} \left(\bar{A}_{k,j}\bar{G}_{k,j}\bar{G}_{k,j}^\top \bar{A}_{k,j}^\top + \bar{G}_{k,j+1}\bar{G}_{k,j+1}^\top\right)
\end{align}
where $\Delta t_{k,j} = t_{k,j+1} - t_{k,j}$, $\bar{G}_{k,j} = G_x(\bar{\bm{x}}_{k,j}, \bar{\bm{u}}_{k})$, and $\bar{G}_{k,j+1} = G_x(\bar{\bm{x}}_{k,j+1}, \bar{\bm{u}}_{k})$.
Accordingly, $\bar{G}_{x_{k,j}}$ can be written as
\begin{align}
    \bar{G}_{x_{k,j}} = \sqrt{\frac{\Delta t_{k,j}}{2}}
    \begin{bmatrix}
        \bar{A}_{k,j}\bar{G}_{k,j} & \bar{G}_{k,j+1}
    \end{bmatrix}
\end{align}
so that $\bar{Q}_{k,j} = \bar{G}_{x_{k,j}} \bar{G}_{x_{k,j}}^\top$.
This approximation is significantly faster than numerically integrating \Eqref{eq:lyapunov} and more accurate than the Euler approximation used in Refs.~\cite{Theodorou2010-xy, Yi2020-mz}.

\subsection{Representative cost functions and constraints}
Since the covariance matrices $\tilde{P}_k$ and $\hat{P}_k$, together with the linear feedback gain $K_k$, are included in the state and control vectors, the costs and constraints can be formulated explicitly as functions of these quantities.
In other words, the proposed method can handle stochastic costs and chance constraints expressed in terms of moments up to second order.
Here, we introduce representative cost functions and constraints commonly used in spacecraft trajectory design, including minimum-fuel costs with covariance penalties, a stochastic thrust magnitude constraint, terminal covariance constraints, and linear state chance constraints.
The cost functions and constraints introduced here are twice continuously differentiable with respect to the reference trajectory $\left(\bar{\bm{X}}_k, \bar{\bm{U}}_k\right)$ and can therefore be incorporated into the proposed method.
If different cost functions or constraints are introduced, their derivatives can also be evaluated using AD, although analytic implementations are significantly faster in practice.

\subsubsection{Cost functions}
The stage and terminal cost functions are defined as the sum of the smoothed $\ell_2$-norm of the nominal control and quadratic penalties on the deviations of the state and control from the nominal trajectory.
The cost functions are written as
\begin{subequations}
\begin{align} \label{eq:typical_cost}
    L_k(\bm{X}_k, \bm{U}_k) &= \Delta t_k \left(\sqrt{\lVert \bar{\bm{u}}_k\rVert ^2+\epsilon_u}  + \mathrm{tr} \left[\hat{P}_k Q_k \right] + \mathrm{tr} \left[\tilde{P}_k Q_k \right] + \mathrm{tr}\left[P_{u_k} R_k \right] \right)\\
    \varphi(\bm{X}_N) &= \mathrm{tr} \left[\hat{P}_N Q_N \right] + \mathrm{tr} \left[\tilde{P}_N Q_N \right] \label{eq:typical_cost_t}
\end{align}
\end{subequations}
where $\Delta t_k := t_{k+1} - t_k$ is the discretization time step, $\epsilon_u$ is a small mass-leak parameter, $Q_k \succeq O_{n_x}$ and $R_k \succ O_{n_u}$ are weighting matrices, and $P_{u_k} := P_{u_{k,0}} = K_k\hat{P}_kK_k^\top$; thus, the stochastic control penalty is evaluated only at the maneuver epochs.

The optimization results depend on the choice of the weighting matrices.
With small weighting matrices, the solution is nearly fuel-optimal.
In contrast, with large weighting matrices, the resulting solution becomes more observability-aware or uncertainty-robust: the optimized reference trajectory passes through a highly observable region, or the feedback gain keeps the spacecraft close to the reference trajectory, thereby reducing the covariances.

\subsubsection{Thrust magnitude constraint} \label{sec:thrust_mag_con}
The chance constraint on the thrust magnitude is given by
\begin{align} \label{eq:control_chance_const}
    \mathbb{P}\left[\lVert \bm{u}_k\rVert  \leq u_{\mathrm{max}} \right] \geq 1 - \varepsilon_u
\end{align}
where $\bm{u}_k := \bm{u}_{k,0}$, $u_{\mathrm{max}}$ is the maximum thrust magnitude, and $\varepsilon_u$ is a risk bound.
The sufficient condition derived in Ref.~\cite{Oguri_undated-kc} has the following convex form:
\begin{align} \label{eq:thrust_mag_convex}
    \lVert \bar{\bm{u}}_k\rVert  + n_{\sigma}(\varepsilon_u, n_u) \sqrt{\lVert P_{u_k} \rVert} \leq u_{\mathrm{max}}
\end{align}
where $\lVert P_{u_k} \rVert$ is the spectral norm of $P_{u_k}$, $n_\sigma (\varepsilon_u, n_u) = \sqrt{\mathcal{Q}_{X \sim \chi^2(n_u)}(1-\varepsilon_u)}$, and $\mathcal{Q}_{X \sim \chi^2(n_u)}(1-\varepsilon)$ denotes the quantile function of the chi-squared distribution with $n_u$ degrees of freedom evaluated at probability $1-\varepsilon$.

However, this convex formulation is not straightforward to handle in the PO-SDDP algorithm because the derivatives of the first term are singular when $\lVert \bar{\bm{u}}_k\rVert = 0$, and the derivatives of $\lVert P_{u_k} \rVert$ are also ill-conditioned when $P_{u_k}$ has repeated eigenvalues.
We approximate \Eqref{eq:thrust_mag_convex} using a smooth function based on the Schatten $p$-norm:
\begin{align} \label{eq:approx_schatten}
    \sqrt{\lVert \bar{\bm{u}}_k\rVert ^2+\epsilon_u} + n_{\sigma}(\varepsilon_u, n_u) \sqrt{\left[\mathrm{tr}(P_{u_k}^p)\right]^{1/p}} \leq u_{\mathrm{max}}
\end{align}
where the first term is regularized by the mass-leak parameter also used in \Eqref{eq:typical_cost}, and, in the second term, the spectral norm in \Eqref{eq:thrust_mag_convex} is replaced by the Schatten $p$-norm.
In this constraint, $P_{u_k}$ is also evaluated at the maneuver epoch, i.e., $P_{u_k}=P_{u_{k,0}}$.
This Schatten surrogate overestimates $\sqrt{\lVert P_{u_k} \rVert}$ by a factor of $n_u^{1/2p}$ in the worst case.

\subsubsection{Terminal covariance constraint}
The terminal covariance constraint is generally formulated as the matrix inequality
\begin{align} \label{eq:terminal_cov_con}
    P_N \approx \tilde{P}_N + \hat{P}_N \preceq P_f
\end{align}
where $P_f$ represents the target state covariance.

The scalar inequality equivalent to \Eqref{eq:terminal_cov_con} can be written as
\begin{align} \label{eq:terminal_guidance_error}
    \lVert \mathcal{S}_N \rVert - 1 \leq 0, & & \mathcal{S}_N = P_f^{-\frac{1}{2}} \left(\tilde{P}_N + \hat{P}_N\right) P_f^{-\frac{1}{2}}
\end{align}
As mentioned in \Secref{sec:thrust_mag_con}, the derivatives of the spectral norm are singular when the matrix $\mathcal{S}_N$ has repeated eigenvalues.
For numerical stability, the logarithm of the normalized Schatten $p$-norm of $\mathcal{S}_N$ is introduced here:
\begin{align} \label{eq:terminal_cov_con_approx}
    \frac{1}{p}\log{\left[ \frac{\mathrm{tr}\left(\mathcal{S}_N^p\right)}{n_x}\right]} \leq 0
\end{align}
This surrogate can underestimate $\lVert \mathcal{S}_N \rVert$ by a factor of $1/n_x^{1/p}$ in the worst case.

\subsubsection{Linear state chance constraint}
A linear state chance constraint can be formulated as
\begin{align} \label{eq:lin_state_cc}
    \mathbb{P}\left[\bm{a}^\top_s \bm{x}_k \leq b_s \right] \geq 1 - \varepsilon_x
\end{align}
where $\bm{a}_s$ and $b_s$ define a feasible half-space in which the spacecraft can move, and $\varepsilon_x$ is a risk bound.
Under a Gaussian belief, the deterministic constraint that is necessary and sufficient for \Eqref{eq:lin_state_cc} can be expressed as
\begin{align} \label{eq:state_chance_con}
    \bm{a}^\top_s  \bar{\bm{x}}_k
    + \Psi^{-1} (1-\varepsilon_x)\,
    \left\lVert \left(\tilde{P}_k + \hat{P}_k \right)^{\frac{1}{2}} \bm{a}_s \right\rVert
    - b_s
    \leq 0
\end{align}
where $\Psi^{-1} (1-\varepsilon_x)$ denotes the inverse cumulative distribution function at probability $1-\varepsilon_x$.
This constraint can be handled in the PO-SDDP algorithm without approximation.

\section{Numerical Examples} \label{sec:numerical_examples}
In this section, the PO-SDDP algorithm is demonstrated through three scenarios: the light-dark domain problem, the Earth-to-Mars transfer problem, and the halo orbit transfer in the Earth--Moon CR3BP\@.
All computations are performed on a desktop computer (Intel Core i9-14900KF, 3.2 GHz).

In all scenarios, the stage and terminal cost functions are defined by \Eqref{eq:typical_cost} and \Eqref{eq:typical_cost_t}, respectively.
The initial augmented controls are set to zero, i.e., $\bm{U}_k = \bm{0}_{n_U}$ for all $k$, and the initial Lagrange multipliers for all constraints are also initialized with zero vectors.
The \texttt{Vern7} integrator from \texttt{DifferentialEquations.jl}~\cite{rackauckas2017differentialequations} is used to discretize the continuous dynamics, except for the light-dark domain problem, where the dynamics are linear.

To validate the optimization results, Monte Carlo analysis is conducted using 500 random initial states sampled from $\mathcal{N}(\bm{x}_0, \ \tilde{P}_0 + \hat{P}_0)$, with dynamical disturbances and random observations.
For each sampled trajectory, the state is estimated using the standard EKF and controlled by the optimal policy given in \Eqref{eq:controller}.
In the Monte Carlo analysis, the process-noise covariance is computed by numerically integrating the Lyapunov equation in \Eqref{eq:lyapunov} to assess the accuracy of the trapezoidal approximation used in the optimization.

\subsection{Light-dark domain problem} \label{sec:ex1}
Here, we consider the light-dark domain problem in which observation uncertainty varies with distance from a landmark.
The state is defined by the two-dimensional position and velocity, $\bm{x} = [ r_x \ \  r_y  \ \  v_x \ \  v_y ]^\top$, and the dynamics are modeled as a double integrator described by the following discrete-time linear system:
\begin{align}
    \bm{x}_{k+1} = A_k \bm{x}_k + B_k \bm{u}_k + G_{x_k} \bm{w}_x
\end{align}
with system matrices
\begin{align} \label{eq:dyn_ex1}
    A_k =
    \begin{bmatrix}
        1 & 0 & \Delta t_k & 0 \\
        0 & 1 & 0 & \Delta t_k \\
        0 & 0 & 1 & 0 \\
        0 & 0 & 0 & 1
    \end{bmatrix}, & &
    B_k =
    \begin{bmatrix}
        0.5 \Delta t_k^2 & 0 \\
        0 & 0.5 \Delta t_k^2 \\
        \Delta t_k & 0 \\
        0 & \Delta t_k
    \end{bmatrix}, & &
    G_{x_k} =
    \begin{bmatrix}
        \sigma_{p} & 0 & 0 & 0 \\
        0 & \sigma_{p} & 0 & 0 \\
        0 & 0 & \sigma_{v} & 0 \\
        0 & 0 & 0 & \sigma_{v}
    \end{bmatrix}
\end{align}
where $\sigma_{p}$ and $\sigma_{v}$ are positive scalars representing the disturbance intensities of the position and velocity, respectively.

The observation is defined as the position with noise proportional to the distance from the landmark:
\begin{align}
    \bm{y}_{k,j} = \bar{C}_{k,j} \bm{x}_{k,j} + G_{y} \bm{w}_y
\end{align}
with
\begin{align}
    \bar{C}_{k,j} =
    \begin{bmatrix}
        1 & 0 & 0 & 0 \\
        0 & 1 & 0 & 0
    \end{bmatrix}, & &
    G_y =
    \begin{bmatrix}
        \sigma_{y,0} + r_{k,j} \sigma_{y,1} & 0 \\
        0 & \sigma_{y,0} + r_{k,j} \sigma_{y,1}
    \end{bmatrix}, & &
    r_{k,j} = \sqrt{(r_{x_{k,j}} - l_x)^2 + (r_{y_{k,j}} - l_y)^2}
\end{align}
where $\sigma_{y,0}$ is a fixed observation-noise term and $\sigma_{y,1}$ is a coefficient for the noise proportional to the distance from the landmark position $[ l_x  \ \ l_y ]^\top$.

The key parameters in this scenario are summarized in \Tabref{tab:param1}.
For the cost function, the weighting matrices are set to $Q_k = O_{n_x}$ and $R_k = I_{n_u}$, the scaling matrix in the trust-region subproblem is chosen as the identity matrix, i.e., $D_{\mathrm{tr}} = I_{n_U}$, and the mass-leak parameter is set to $\epsilon_u = 10^{-8}$.
The stochastic thrust magnitude, terminal covariance, and linear state constraints are imposed, together with the terminal nominal-state constraint $\bar{\bm{x}}_N = \bm{x}_f$ where $\bm{x}_f$ denotes the target nominal state.
The initial estimation-error and state-estimate covariances are both specified by the standard deviations $\tilde{\sigma}_{r_0}=\hat{\sigma}_{r_0}=4\times10^{-2}$ and $\tilde{\sigma}_{v_0}=\hat{\sigma}_{v_0}=10^{-2}$, while the terminal target covariance is specified by $\sigma_{r_f}=2\times10^{-4}$ and $\sigma_{v_f}=10^{-2}$.
For the augmented Lagrangian method, the initial penalties are set to $1.0$ for all constraints, the penalty scaling factor is set to $2.0$, and the Schatten-norm parameters are chosen as $p=1$ for the thrust magnitude constraint and $p=8$ for the surrogate terminal covariance constraint.

\begin{table}[tb]
    \caption{Parameters in the light-dark domain scenario}
    \label{tab:param1}
    \centering
    \begin{tabular}{lll}
        \hline
        Parameter & Symbol & Value \\
        \hline
        Number of stages & $N$ & 50 \\
        Time step & $\Delta t_{k,j}, \forall k,j$ & $0.2$ \\
        Number of observations per stage & $N_j(k), \forall k$ & $1$ \\
        Initial nominal state & $\bar{\bm{x}}_0$ & $[0 \ \ 0 \ \ 0 \ \ 0]^\top$ \\
        Target nominal state & $\bm{x}_f$ & $[10 \ \ 0 \ \ 0 \ \ 0]^\top$ \\
        Position noise coefficient & $\sigma_p$ & $10^{-6}$ \\
        Velocity noise coefficient & $\sigma_v$ & $10^{-6}$ \\
        Landmark position & $[l_x,\ l_y]$ & $[5,\ 5]$ \\
        Fixed observation noise & $\sigma_{y,0}$ & $10^{-4}$ \\
        Distance-dependent noise coefficient & $\sigma_{y,1}$ & $10^{-2}$ \\
        Risk bound of state & $\varepsilon_x$ & $10^{-3}$ \\
        Risk bound of control & $\varepsilon_u$ & $10^{-3}$ \\
        Maximum acceleration & $u_{\mathrm{max}}$ & $2.0$ \\
        Half-space coefficient vector & $\bm{a}_s$ & $[0 \ \ 1 \ \ 0 \ \ 0]^\top$ \\
        Half-space bound & $b_s$ & $3$ \\
        \hline
    \end{tabular}
\end{table}

\begin{figure}[!tb]
    \centering
    \includegraphics[width=\textwidth]{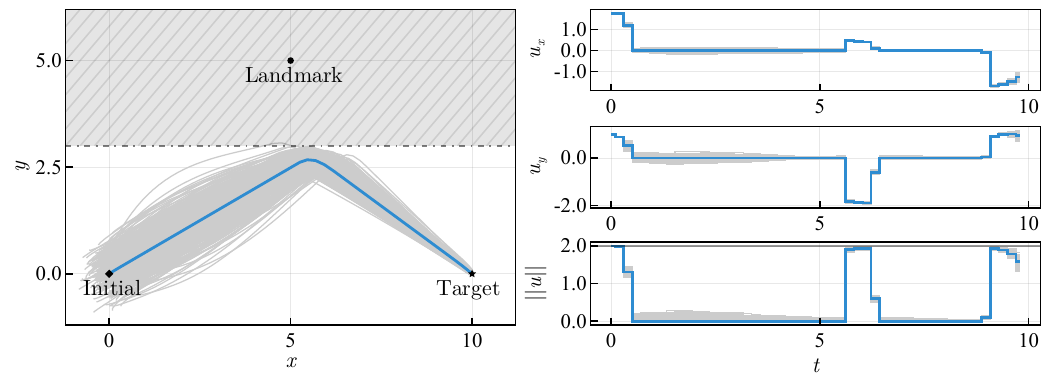}
    \caption{Nominal and sampled trajectories and control profiles in the light-dark domain problem} \label{fig:results_proposed}
\end{figure}

The nominal and sampled trajectories are shown in \Figref{fig:results_proposed}, where the nominal trajectory is plotted as a colored line, and the sampled trajectories are shown in gray. The nominal and sampled control profiles are also presented in \Figref{fig:results_proposed}.
The region defined by $y>3$ in the left panel of \Figref{fig:results_proposed} is designated as a keep-out zone.
Under the obtained control policy, the nominal trajectory approaches the landmark while proceeding toward the target state to obtain lower-noise position measurements. The trajectory then changes direction slightly below $y=3$, leaving a margin to account for process noise and estimation error. The Monte Carlo sample trajectories also avoid entering the keep-out zone.
For the thrust magnitude constraint, the nominal control sequence remains below the specified maximum acceleration, with a larger margin near the final stage.
This increased margin helps reduce the terminal covariance because the system moves away from the landmark near the end of the trajectory, which increases the observation noise.

To clarify how the algorithmic formulation affects the solution, we compare PO-SDDP with belief-space iterative linear quadratic Gaussian (iLQG)~\cite{Van_Den_Berg2012-fn} and the SCP-based approach. Belief-space iLQG corresponds to a variant of the proposed method in which the state-estimate covariance is excluded from the state variables and the second-order STMs for the augmented dynamics are omitted. The SCP-based approach is implemented according to Ref.~\cite{Kumagai2025-bl} using \texttt{JuMP.jl}~\cite{Lubin2023} for general optimization modeling and \texttt{Clarabel.jl}~\cite{Clarabel_2024} as a semidefinite programming solver.
The constraints and cost functions are selected to be consistent with the respective formulations of the methods being compared. In belief-space iLQG, the thrust magnitude, terminal covariance, and linear state constraints are imposed only for the nominal state and the estimation-error covariance. The cost function is defined without the state-estimate covariance term, since it is not included in the state variables. By contrast, the SCP-based approach can handle convex constraints expressed in terms of the spectral norm; therefore, Eqs.~\eqref{eq:thrust_mag_convex} and \eqref{eq:terminal_cov_con} are employed for the thrust magnitude and terminal covariance constraints, respectively. The cost function for the SCP-based approach follows the formulation in Ref.~\cite{Kumagai2025-bl}.
However, with the same target covariance used for PO-SDDP and belief-space iLQG, the adopted SCP-based formulation does not yield a feasible converged solution satisfying the terminal covariance constraint. Therefore, the target covariance for the SCP-based approach is relaxed to $2.5P_f$.

\begin{figure}[!tb]
    \centering
    \includegraphics[width=\textwidth]{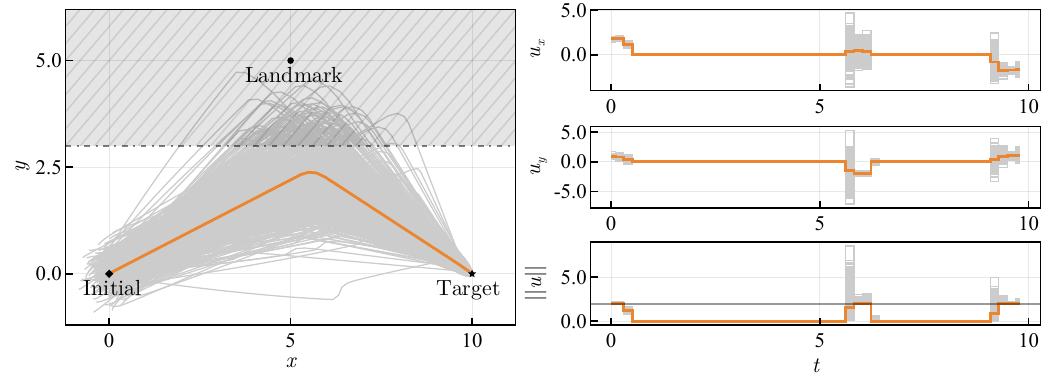}
    \caption{Nominal and sampled trajectories and control profiles obtained by belief-space iLQG} \label{fig:results_belief_ilqg}
\end{figure}

\begin{figure}[!tb]
    \centering
    \includegraphics[width=\textwidth]{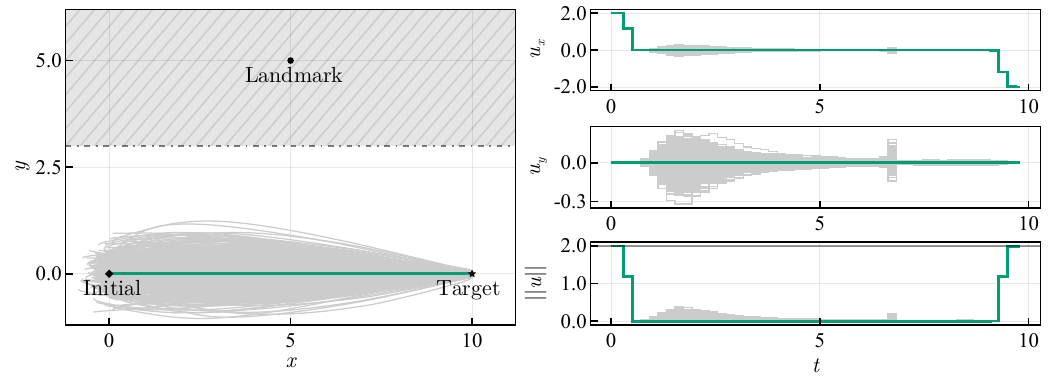}
    \caption{Nominal and sampled trajectories and control profiles obtained by the SCP-based method} \label{fig:results_convex}
\end{figure}

Figures~\ref{fig:results_belief_ilqg} and~\ref{fig:results_convex} show the nominal and sampled trajectories obtained by belief-space iLQG and the SCP-based method, along with the corresponding control sequences.
In the Monte Carlo analysis of belief-space iLQG, the feedback policy obtained in the backward pass is applied to the estimated state. The nominal trajectory generated by belief-space iLQG approaches the landmark in the same manner as that of the proposed method in order to reduce the estimation error. Because the state-estimate covariance and feedback gain are not included as optimization variables, the sampled trajectories and control sequences violate the keep-out zone and thrust magnitude constraints. As shown in \Figref{fig:results_convex}, the SCP-based method yields a minimum-fuel solution with a robust feedback gain that guides the system directly to the target state while controlling the covariance to satisfy the terminal constraints.

Table~\ref{tab:ld_comp} compares the fuel consumption, terminal covariance metric, and computational cost. The proposed method yields a navigation-aware trajectory while maintaining consistency between the predicted and Monte Carlo terminal covariance metrics. In contrast, belief-space iLQG underestimates the terminal uncertainty because the state-estimate covariance and optimized feedback gain are not included in its prediction. The SCP-based method obtains the lowest nominal fuel consumption under a relaxed covariance target, but it does not steer the trajectory toward the landmark because the EKF-related matrices are fixed during convexification. The lower runtime of belief-space iLQG mainly reflects its reduced state dimension.

\begin{table}[!tb]
    \caption{Comparison of the solutions for the light-dark domain scenario}
    \label{tab:ld_comp}
            \centering
    \begin{tabular}{lccccc}
        \hline
        Method & Nominal $\Delta V$, n.d. &  $\lVert\mathcal{S}_N\rVert$ (pred.) & $\lVert\mathcal{S}_N\rVert$ (MC) & Iterations & Runtime, s  \\
        \hline
        PO-SDDP & 3.87 & $1.0904$ & $1.1707$ & 2123 & 29 \\
        Belief-space iLQG & 3.60 & $1.0905$ & $2.0749$ & 1106 & 3.8  \\
        SCP-based & 2.12 & $2.4997$ & $2.8162$ & 80 & 26  \\
        \hline
    \end{tabular}
\end{table}

\subsection{Earth-to-Mars transfer} \label{sec:ex2}
Next, the PO-SDDP algorithm is applied to the design of a low-thrust Earth-to-Mars transfer with radiometric tracking.
The state consists of the spacecraft position $\bm{r}\in\mathbb{R}^2$ and velocity $\bm{v}\in\mathbb{R}^2$ in the heliocentric inertial frame, i.e.,
$\bm{x} = [\bm{r}^\top \ \ \bm{v}^\top]^\top$.
The deterministic part of the continuous-time dynamics is given by
\begin{align}
    \bm{f}_{\mathrm{tb}} =
    \begin{bmatrix}
        \bm{v} \\
        -\mu_s \bm{r}/\lVert \bm{r}\rVert^3 + \bm{u}
    \end{bmatrix}
\end{align}
where $\mu_s$ is the gravitational parameter of the Sun.

The process noise is modeled using a two-dimensional reformulation of the Gates error model~\cite{Gates1963-dv}.
In the original Gates model, the shutoff and resolution errors contribute to the uncertainty along the commanded thrust direction, whereas the pointing and autopilot errors contribute to the uncertainty orthogonal to that direction.
The corresponding noise weighting matrix $G_x(\bm{x}, \bm{u})$ is defined as
\begin{align} \label{eq:gates_ini}
    G_x(\bm{x}, \bm{u}) &= T_{\mathrm{u}} M_{\mathrm{sqrt}}(\bm{u}), &
    T_{\mathrm{u}} =
    \begin{bmatrix}
        O_2 & I_2
    \end{bmatrix}^\top
\end{align}
The square-root factor of the covariance is written as
\begin{align}  \label{eq:gates_end}
    M_{\mathrm{sqrt}}
    =
    \sigma_{\perp} I_2
    +
    \left(\sigma_{\parallel} - \sigma_{\perp}\right)
    \hat{\bm{z}}\hat{\bm{z}}^\top
\end{align}
where the standard deviations orthogonal and parallel to the commanded thrust direction are defined by
\begin{align}
    \sigma_{\perp}
    &=
    \sqrt{\sigma_{\mathrm{ap}}^2 + \sigma_{\mathrm{pt}}^2 \lVert \bm{u} \rVert^2}, &
    \sigma_{\parallel}
    &=
    \sqrt{\sigma_{\mathrm{res}}^2 + \sigma_{\mathrm{sh}}^2 \lVert \bm{u} \rVert^2}
\end{align}
Here, $\sigma_{\mathrm{ap}}$, $\sigma_{\mathrm{pt}}$, $\sigma_{\mathrm{res}}$, and $\sigma_{\mathrm{sh}}$ denote the autopilot, pointing, resolution, and shutoff errors, respectively.
The commanded thrust direction is regularized as
\begin{align}
    \hat{\bm{z}}
    =
    \frac{\bm{u}}{\sqrt{\lVert \bm{u} \rVert^2 + \epsilon_u}}
\end{align}
For $\lVert\bm{u}\rVert^2 \gg \epsilon_u$, this representation yields standard deviations approximately equal to $\sigma_{\parallel}$ along the commanded thrust direction and $\sigma_{\perp}$ in the orthogonal direction.
The regularized direction vector preserves the directional structure of the original Gates model while keeping the noise model differentiable near zero thrust.

Range and range-rate observations between the spacecraft and Earth are modeled as
\begin{align} \label{eq:range_rangerate1}
    \bm{y}_{k,j} =
    \begin{bmatrix}
        \lVert \bm{\rho}_r \rVert \\
        \left(\bm{\rho}_r \cdot \bm{\rho}_v\right)/\lVert \bm{\rho}_r \rVert
    \end{bmatrix}
    +
    \begin{bmatrix}
        \sigma_{y,r} & 0 \\
        0 & \sigma_{y,v}
    \end{bmatrix}\bm{w}_y
\end{align}
where $\bm{\rho}_r = \bm{r} - \bm{r}_{\mathrm{e}}$, $\bm{\rho}_v = \bm{v} - \bm{v}_{\mathrm{e}}$, $\bm{r}_{\mathrm{e}}$ and $\bm{v}_{\mathrm{e}}$ denote the Earth's position and velocity, respectively.

The initial and target states are listed in \Tabref{tab:states2}, and the key parameters for this scenario are summarized in \Tabref{tab:param2}.
The transfer duration and the initial and target states are based on those used in the numerical examples of Refs.~[\citenum{Ozaki2020-id,ridderhof2020chance}], except that the initial state is slightly modified to avoid a singularity in the observation model.
In this scenario, the PO-SDDP algorithm is applied with three different shutoff-error values to compare the resulting solutions.
To demonstrate that the proposed method can accommodate multiple observations, range and range-rate measurements are assumed to be available at three equally spaced epochs in each stage, including the subsequent maneuver epoch. Thus, the $N=40$ maneuver sequence contains 120 intermediate belief-state transitions.

For the cost function, the weighting matrices are set to $Q_k = O_{n_x}$ and $R_k = I_{n_u}$, and the mass-leak parameter is set to $\epsilon_u = 10^{-8}~(\mathrm{mm/s^2})^2$.
The thrust magnitude constraint is imposed together with terminal constraints on the nominal state and terminal covariance.
The initial estimation-error and state-estimate covariances are both specified by the standard deviations  $\tilde{\sigma}_{r_0}=\hat{\sigma}_{r_0}=10~\mathrm{km}$ and $\tilde{\sigma}_{v_0}=\hat{\sigma}_{v_0}=0.1~\mathrm{km/s}$, while the terminal target covariance is specified by $\sigma_{r_f}=3.12\times10^5~\mathrm{km}$ and $\sigma_{v_f}=0.1~\mathrm{km/s}$.
For the augmented Lagrangian method, the initial penalties are set to 1.0 for the thrust magnitude and terminal-state constraints and to $1 \times 10^{-6}$ for the surrogate terminal covariance constraint, while the penalty scaling factor is set to 2.0 for all constraints.
The Schatten-norm parameters are chosen as $p=1$ for the thrust magnitude constraint and $p=8$ for the surrogate terminal covariance constraint.
In the trust-region subproblem, the scaling matrix is set to $D_{\rm tr} = \mathrm{blkdiag}(I_{n_u}, 0.1I_{n_un_x})$, so that the terms corresponding to the feedback gain are one order of magnitude smaller than those corresponding to the nominal control.
Using the identity scaling matrix fails to yield a solution that satisfies the surrogate terminal covariance constraints because the updates to the nominal control and feedback gain are treated as having comparable scales.
The physical parameters are nondimensionalized using the length and time-scale factors listed in \Tabref{tab:param2} to improve numerical stability during the optimization.

\begin{table}[!tb]
    \caption{Initial and target states for the Earth-to-Mars transfer scenario}
    \label{tab:states2}
    \centering
    \begin{tabular}{lcccc}
        \hline
        State & $x$, AU & $y$, AU & $v_x$, km/s & $v_y$, km/s \\
        \hline
        Initial & $-0.94048$ & $-0.34502$ & $9.7746$ & $-29.078$ \\
        Target  & $-1.1543$  & $1.1829$ & $-16.427$ & $-14.861$ \\
        \hline
    \end{tabular}
\end{table}

\begin{table}[!tb]
    \caption{Parameters for the Earth-to-Mars transfer scenario}
    \label{tab:param2}
    \centering
    \begin{tabular}{lccc}
        \hline
        Parameter & Symbol & Value & Unit \\
        \hline
        Number of stages & $N$ & $40$ & - \\
        Transfer duration & $t_N - t_0$ & $348.79$ & days \\
        Gravitational parameter & $\mu_s$ & $1.3271 \times 10^{11}$ & $\mathrm{km^3/s^2}$ \\
        Number of observations per stage & $N_j(k), \ \forall k$ & $3$ & - \\
        Time step & $\Delta t_{k,j}, \ \forall k,j$ & $2.9066$ & days \\
        Autopilot error & $\sigma_{\mathrm{ap}}$ & $1\times10^{-3}$ & $\mathrm{mm/s^2}$  \\
        Pointing error & $\sigma_{\mathrm{pt}}$ & $0.5$ & deg \\
        Resolution error & $\sigma_{\mathrm{res}}$ & $1\times10^{-3}$ & $\mathrm{mm/s^2}$ \\
        Shutoff error & $\sigma_{\mathrm{sh}}$ & $1.0$ (low), $3.0$ (medium), $4.0$ (high) & \% \\
        Range noise & $\sigma_{y,r}$ & $10^2$ & $\mathrm{km}$ \\
        Range-rate noise & $\sigma_{y,v}$ & $1$ & $\mathrm{m/s}$ \\
        Risk bound & $\varepsilon_u$ & $10^{-3}$ & - \\
        Maximum acceleration & $u_{\mathrm{max}}$ & $1.0$ & $\mathrm{mm/s^2}$ \\
        Length scale factor & $L_{\mathrm{sf}}$ & $10^8$ & km \\
        Time scale factor & $T_{\mathrm{sf}}$ & $10^7$ & sec \\
        \hline
    \end{tabular}
\end{table}

\begin{figure}[!tb]
    \centering
    \includegraphics[width=\textwidth]{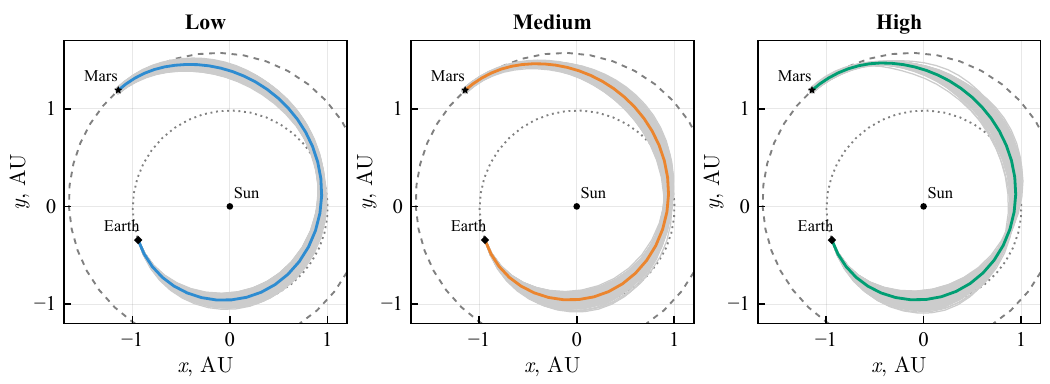}
    \caption{Nominal and sampled trajectories for different shutoff errors, with deviations from the corresponding nominal trajectories exaggerated by a factor of 5} \label{fig:results_etm_traj_comp}
\end{figure}

\begin{figure}[!tb]
    \centering
    \includegraphics[width=\textwidth]{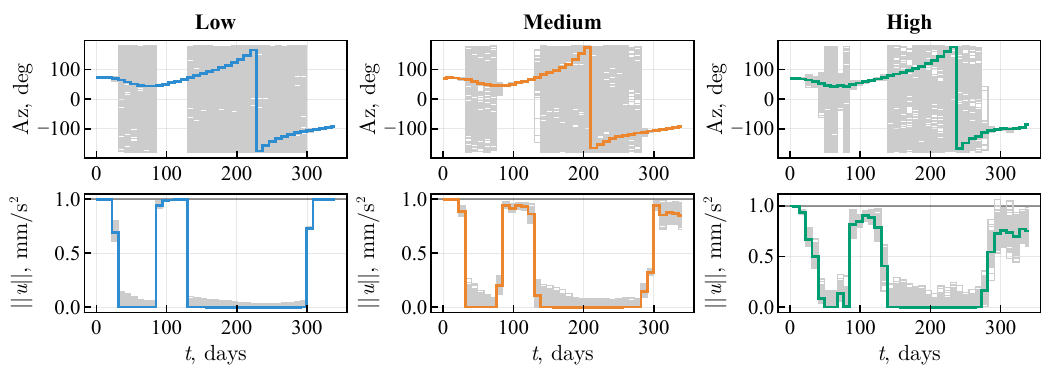}
    \caption{Nominal and sampled control profiles for different shutoff errors} \label{fig:results_etm_ctrl_comp}
\end{figure}

Figure~\ref{fig:results_etm_traj_comp} shows the nominal and sampled trajectories, with the deviations from the nominal trajectories exaggerated.
The nominal control profiles and the sampled profiles from the Monte Carlo analysis are shown in \Figref{fig:results_etm_ctrl_comp}, where the thrust azimuth is defined as $\mathrm{atan2}(u_y, u_x)$.
Figures~\ref{fig:results_etm_traj_comp} and~\ref{fig:results_etm_ctrl_comp} show that the nominal trajectories are nearly identical across the shutoff-error cases, whereas the thrusting durations, thrust margins, and correction maneuvers vary with the maneuver uncertainty.
Thus, in this scenario, the proposed method primarily designs a feedback policy robust to maneuver errors rather than reshaping the nominal trajectory for improved observability.

\begin{figure}[!tb]
    \centering
    \includegraphics[width=\textwidth]{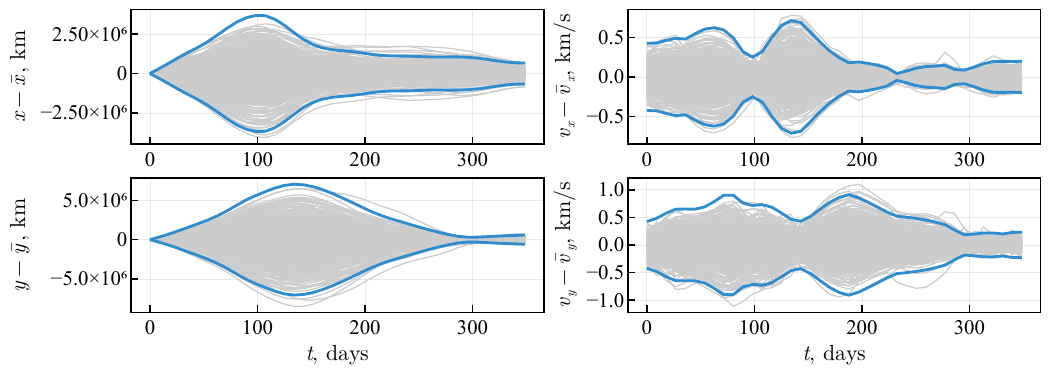}
    \caption{Time histories of position and velocity errors relative to the nominal trajectory for the high-error case, together with the predicted $3\sigma$ envelopes} \label{fig:history_etm_posvel}
\end{figure}

Figure~\ref{fig:history_etm_posvel} shows that the sampled deviations remain mostly within the predicted $3\sigma$ envelopes for the high-error case.
This result supports the accuracy of the trapezoidal approximation and the local covariance propagation used in the optimization.
Table~\ref{tab:e2m_comp} shows that the required $\Delta V$ increases with the shutoff-error level, reflecting the larger control margin required under stronger maneuver uncertainty.

\begin{table}[!tb]
    \caption{Comparison of Earth-to-Mars transfer solutions}
    \label{tab:e2m_comp}
    \centering
    \begin{tabular}{lccccc}
        \hline
        Shutoff-error case & Nominal $\Delta V$, km/s & $\lVert \mathcal{S}_N \rVert$ (pred.) & $\lVert \mathcal{S}_N \rVert$ (MC) & Iterations & Runtime, s \\
        \hline
        Low    & 10.3 & $1.19$ & $1.25$ & 2158 & 72 \\
        Medium & 10.5 & $1.18$ & $1.13$ & 3241 & 99 \\
        High   & 10.9 & $0.69$ & $0.70$ & 3732 & 101 \\
        \hline
    \end{tabular}
\end{table}

\subsection{Halo orbit transfer in Earth--Moon CR3BP} \label{sec:ex3}
As a third demonstration, the PO-SDDP algorithm is applied to an $L_2$-halo-to-$L_1$-halo transfer in the Earth--Moon CR3BP with radiometric tracking.
The state consists of the three-dimensional position $\bm{r} \in \mathbb{R}^3$ and velocity $\bm{v} \in \mathbb{R}^3$ in the rotating frame with the origin at the barycenter of the system, i.e., $\bm{x} = [\bm{r}^\top \ \ \bm{v}^\top]^\top$.
The continuous-time dynamics are given by
\begin{align} \label{dyn:crtbp}
    \bm{f}_{\mathrm{crtbp}}(\bm{x}, \bm{u}) =
    \begin{bmatrix}
        \bm{v} \\
        \bm{a}_{\mathrm{crtbp}}(\bm{r}, \bm{v}) + \bm{u}
    \end{bmatrix}, \quad
    \bm{a}_{\mathrm{crtbp}}(\bm{r}, \bm{v}) =
    \begin{bmatrix}
        2v_y + r_x - \frac{(1-\mu)(r_x+\mu)}{\rho_1^3} - \frac{\mu(r_x-1+\mu)}{\rho_2^3} \\
        -2v_x + r_y - \frac{(1-\mu)r_y}{\rho_1^3} - \frac{\mu r_y}{\rho_2^3} \\
        - \frac{(1-\mu)r_z}{\rho_1^3} - \frac{\mu r_z}{\rho_2^3}
    \end{bmatrix}
\end{align}
where $\mu = m_2 / (m_1 + m_2)$ is the mass parameter of the CR3BP, and $m_1$ and $m_2$ denote the masses of the primary and secondary bodies, respectively.
The relative position vectors to the two primaries are defined as $\bm{\rho}_1 = [r_x+\mu  \ \ r_y \ \  r_z]^\top$ and $\bm{\rho}_2 = [r_x-(1-\mu)  \ \ r_y \ \  r_z ]^\top$ with $\rho_1 = \|\bm{\rho}_1\|$ and $\rho_2 = \|\bm{\rho}_2\|$.
The dynamical noise is modeled by a three-dimensional Gates error model obtained by extending the two-dimensional model described in \Eqref{eq:gates_ini} and \Eqref{eq:gates_end}.
The observations are modeled as range and range-rate measurements from the Earth in the rotating frame, as described in \Eqref{eq:range_rangerate1}, with $\bm{r}_{\mathrm{e}} = [-\mu \ \ 0 \ \ 0]^\top$ and $\bm{v}_{\mathrm{e}} = \bm{0}$.

The initial and target states are taken from Ref.~\cite{Aziz2019-bw}, as listed in \Tabref{tab:states3}, and the key parameters are summarized in \Tabref{tab:param3}.
In this scenario, two state-weighting matrices, $Q_k = 10I_{n_x}$ and $Q_k = 500I_{n_x}$ for all $k$, are considered in the cost function to compare the resulting solutions, while the control-weighting matrix is fixed at $R_k = I_{n_u}$ and the mass-leak parameter is set to $\epsilon_u = 7.4\times 10^{-8}~(\mathrm{mm/s^2})^2$.
For the trust-region method, the scaling matrix in the trust-region subproblem is chosen as the identity matrix, i.e., $D_{\mathrm{tr}} = I_{n_U}$.
Here, the thrust magnitude constraint and the terminal constraint on the nominal state are imposed.
For the augmented Lagrangian method, the initial penalties for the thrust magnitude and terminal-state constraints are both set to $1.0$, and the penalty scaling factor is set to $2.0$ for both constraints.
The Schatten-norm parameter is chosen as $p=1$ for the thrust magnitude constraint.
The initial estimation-error covariance and state-estimate covariance are both defined from the standard deviations $\tilde{\sigma}_r=\hat{\sigma}_r=10~\mathrm{km}$ and $\tilde{\sigma}_v=\hat{\sigma}_v=0.1~\mathrm{m/s}$.

\begin{table}[tb]
    \caption{Initial and target states for the halo orbit transfer scenario}
    \label{tab:states3}
    \centering
    \begin{tabular}{ccccccc}
        \hline
        State & $x$ & $y$ & $z$ & $v_x$ & $v_y$ & $v_z$ \\
        \hline
        Initial & $1.16$ & $0.0$ & $-0.122697$ & $0.0$ & $-0.207128$ & $0.0$ \\
        Target  & $0.85$  & $0.0$ & $0.173890$ & $0.0$ & $0.262114$ & $0.0$ \\
        \hline
    \end{tabular}
\end{table}

\begin{table}[tb]
    \caption{Parameters for the halo-orbit transfer scenario}
    \label{tab:param3}
            \centering
    \begin{tabular}{lccc}
        \hline
        Parameter & Symbol    & Value & Unit\\
        \hline
        Number of stages & $N$  & $120$ & - \\
        Transfer duration & $t_N - t_0$ & $19.1$ & days\\
        Number of observations per stage & $N_j(k), \ \forall k$ & $1$ & - \\
        Time step for control & $\Delta t_{k,j}, \forall k,j$  & $3.8$  & hours \\
        Autopilot error & $\sigma_{\mathrm{ap}}$ & $10^{-3}$ & $\mathrm{mm/s^2}$ \\
        Pointing error & $\sigma_{\mathrm{pt}}$ & $0.5$ & deg \\
        Resolution error & $\sigma_{\mathrm{res}}$ & $10^{-3}$ & $\mathrm{mm/s^2}$ \\
        Shutoff error & $\sigma_{\mathrm{sh}}$ & $1.0$ & \% \\
        Range noise & $\sigma_{y,r}$ & $10^2$ & $\mathrm{km}$ \\
        Range-rate noise & $\sigma_{y,v}$ & $1$ & $\mathrm{m/s}$ \\
        Risk bound & $\varepsilon_u$ & $10^{-3}$ & - \\
        Maximum acceleration & $u_\mathrm{max}$ & $0.75$ & $\mathrm{mm/s^2}$\\
        \hline
    \end{tabular}
\end{table}

\begin{figure}[!tb]
    \centering
    \includegraphics[width=\textwidth]{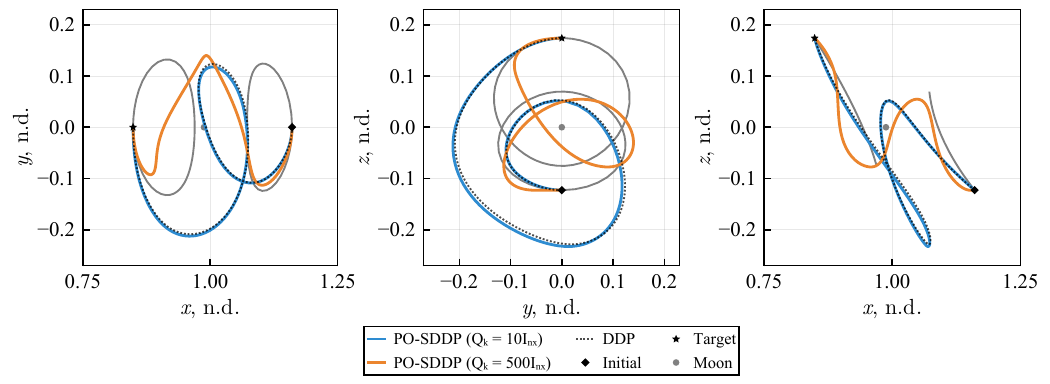}
    \caption{Nominal trajectories projected onto the coordinate planes}
    \label{fig:2d_traj_crtbp}
\end{figure}

\begin{figure}[!tb]
    \centering
    \includegraphics[width=82mm]{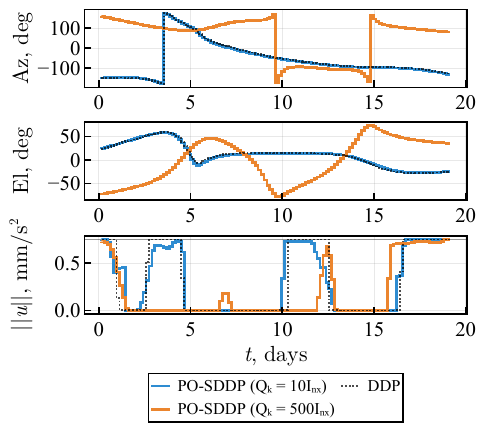}
    \caption{Nominal control profiles for the halo-orbit transfer scenario}
    \label{fig:ctrl_nom_crtbp}
\end{figure}

Figures~\ref{fig:2d_traj_crtbp} and~\ref{fig:ctrl_nom_crtbp} compare the nominal trajectories and control profiles.
The thrust magnitude, azimuth ($\operatorname{atan2}(u_y,u_x)$), and elevation ($\arcsin(u_z/\lVert\bm{u}\rVert)$) histories in \Figref{fig:ctrl_nom_crtbp} show that the solution obtained with the large weighting matrix differs from the other solutions in the timing, magnitude, and three-dimensional directions of the thrust arcs.
With the small weighting matrix, the PO-SDDP solution behaves as a robustified version of the deterministic DDP solution, producing a similar nominal trajectory with more conservative correction capability.
With the large weighting matrix, the optimized trajectory follows a different manifold and requires shorter second and third thrusting arcs, indicating that the covariance-related terms can alter the nominal trajectory when they are weighted sufficiently.

\begin{figure}[!tb]
    \centering

    \begin{subfigure}{0.95\linewidth}
        \centering
        \includegraphics[width=\linewidth]{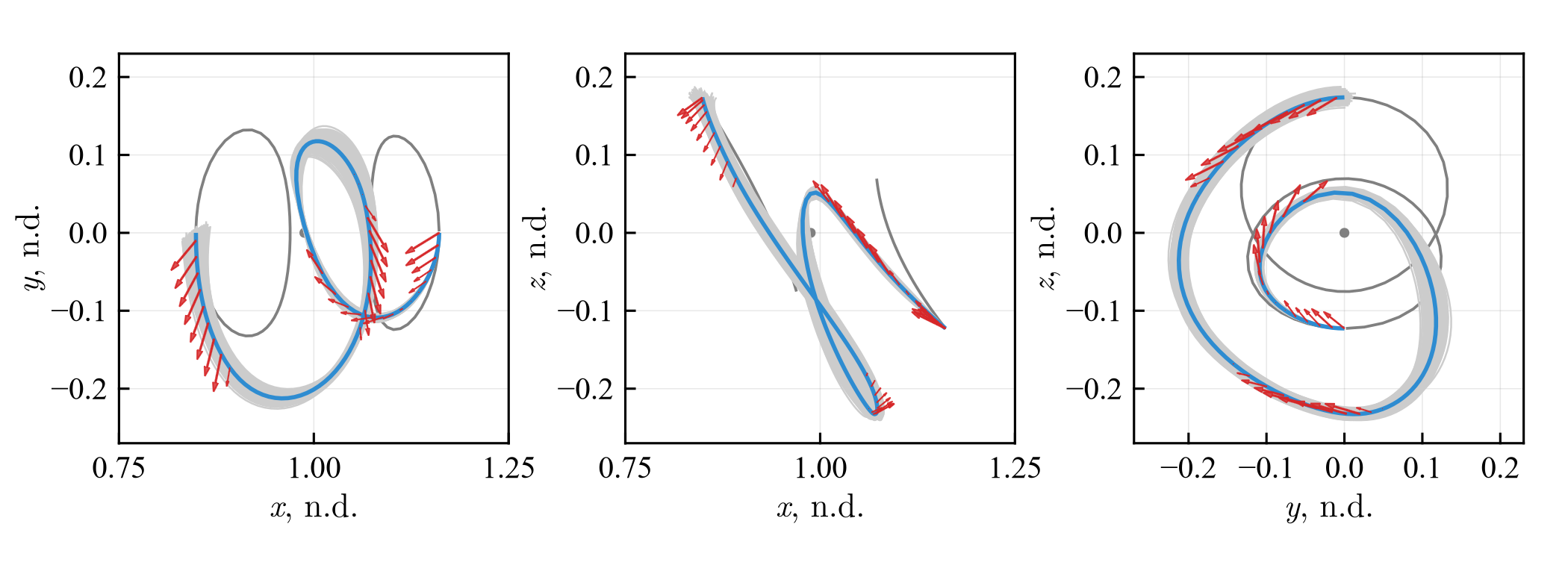}
        \caption{$Q_k = 10I_{n_x}$}
    \end{subfigure}

    \begin{subfigure}{0.95\linewidth}
        \centering
        \includegraphics[width=\linewidth]{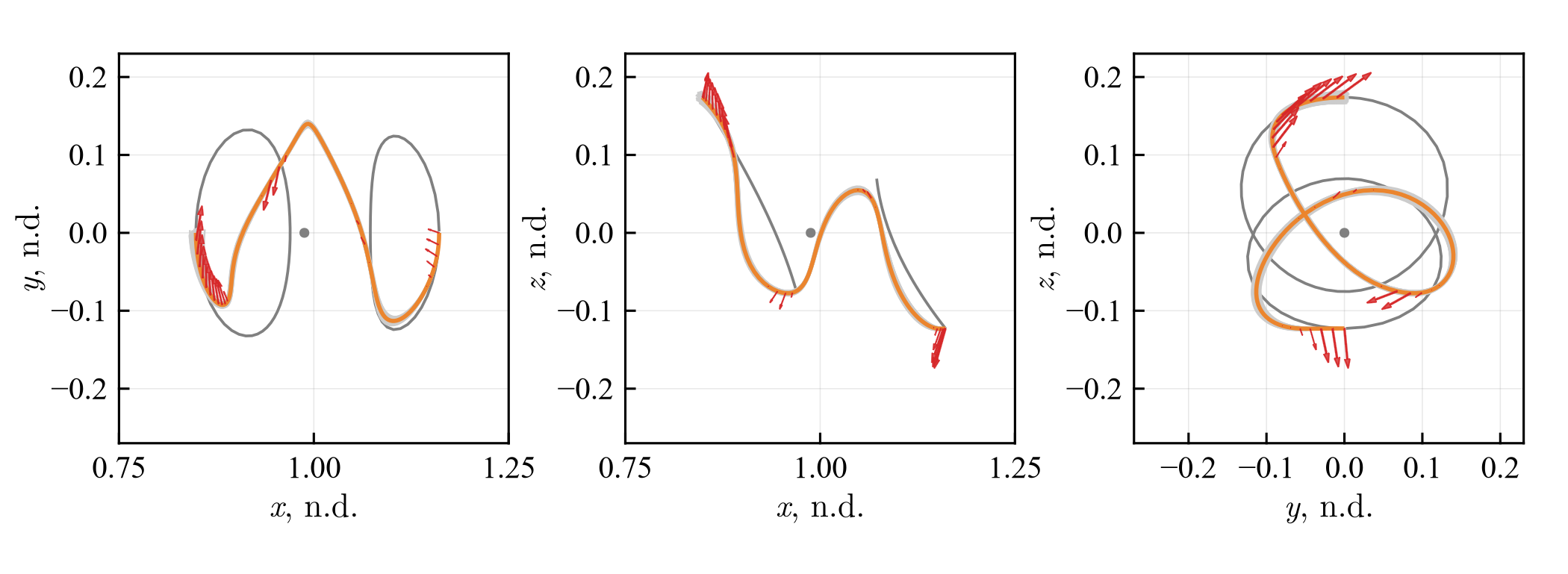}
        \caption{$Q_k = 500I_{n_x}$}
    \end{subfigure}

    \caption{Nominal and sampled trajectories for the halo-orbit transfer scenario, with deviations from the corresponding nominal trajectories exaggerated by a factor of 3}
    \label{fig:3d_traj_crtbp}
\end{figure}

\begin{figure}[!tb]
    \centering
    \includegraphics[width=\textwidth]{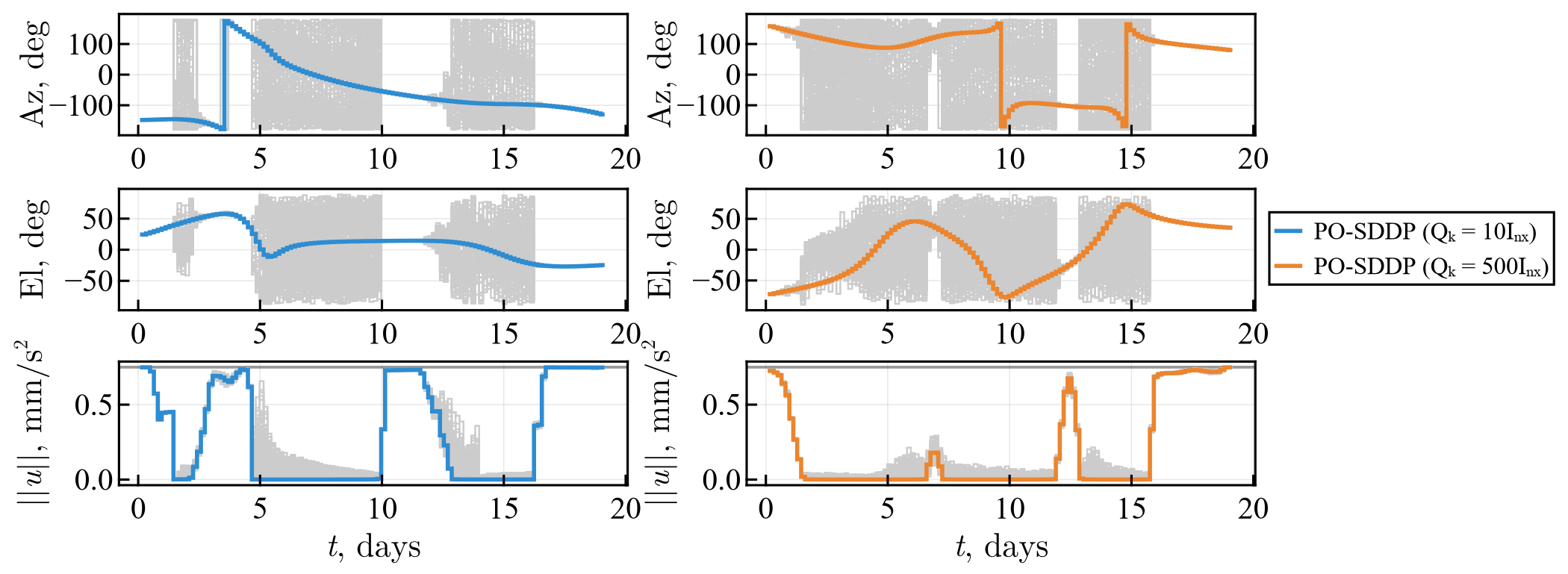}
    \caption{Nominal and sampled control profiles for the halo-orbit transfer solutions with different weighting parameters}
    \label{fig:ctrl_mc_crtbp}
\end{figure}

Figures~\ref{fig:3d_traj_crtbp} and~\ref{fig:ctrl_mc_crtbp} show the corresponding Monte Carlo trajectories and control histories. The large-weighting solution exhibits smaller deviations from the nominal trajectory and requires less corrective maneuvering after the main thrust arcs.

\begin{figure}[!tb]
    \centering
    \includegraphics[width=82mm]{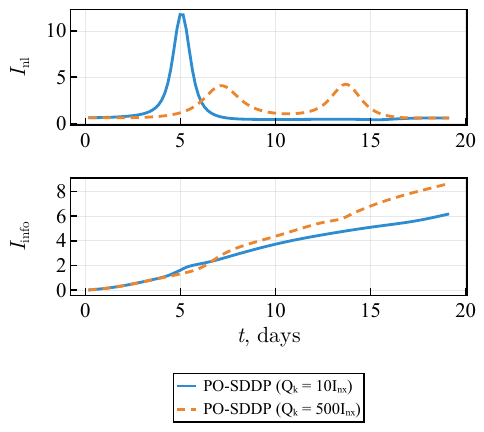}
    \caption{Nonlinearity and information indices along the nominal trajectories}
    \label{fig:ctrl_info_ftle}
\end{figure}

To clarify the differences between the solutions, nonlinearity and information analyses are conducted along the nominal trajectories.
The results are shown in \Figref{fig:ctrl_info_ftle}, where the nonlinearity index is defined using the finite-time Lyapunov exponent (FTLE) as
\begin{align}
    I_{\mathrm{nl},k}
    =
    \frac{1}{\Delta t_k}
    \log \sqrt{\left\|\Phi_A(t_{k+1}, t_k)^\top \Phi_A(t_{k+1}, t_k)\right\|}
\end{align}
and the information index is defined from the cumulative information matrix with respect to the initial state:
\begin{align}
    I_{\mathrm{info},k}
    =
    \log \det \left[
        \tilde{P}_0^{-1}
        +
        \sum_{j=1}^{k}
        \Phi_A(t_j, t_0)^\top C_j^\top W_j C_j \Phi_A(t_j, t_0)
    \right]
\end{align}
where $C_j$ and $W_j$ denote the observation sensitivity matrix and the inverse observation-noise covariance matrix at each stage, respectively.
The nonlinearity index reflects the local sensitivity of the dynamics, whereas the information index quantifies the cumulative information gain with respect to the initial state.
As shown in \Figref{fig:2d_traj_crtbp} and \Figref{fig:ctrl_info_ftle}, the large-weighting solution avoids the strongly nonlinear region near the Moon while maintaining comparable information gain. This behavior explains the reduced terminal dispersion reported in \Tabref{tab:crtbp_comp}. The predicted and Monte Carlo terminal covariance traces are consistent in both PO-SDDP cases.

\begin{table}[!tb]
    \caption{Comparison of the halo-orbit transfer solutions}
    \label{tab:crtbp_comp}
    \centering
    \setlength{\tabcolsep}{3.5pt}
    \renewcommand{\arraystretch}{1.1}
    \begin{tabular}{@{}lccccc@{}}
        \hline
        Method
        & \begin{tabular}{@{}c@{}}Nominal $\Delta V$ \\ km/s\end{tabular}
        & \begin{tabular}{@{}c@{}}$\mathrm{tr}(P_N)$\\pred., n.d.\end{tabular}
        & \begin{tabular}{@{}c@{}}$\mathrm{tr}(P_N)$\\MC, n.d.\end{tabular}
        & Iterations
        & \begin{tabular}{@{}c@{}}Runtime, s\end{tabular} \\
        \hline
        DDP & 0.510 & -- & -- & 4520 & 145 \\
        PO-SDDP, $Q_k=10I_{n_x}$ & 0.511 & $3.94\times10^{-5}$ & $4.32\times10^{-5}$ & 13284 & 2970 \\
        PO-SDDP, $Q_k=500I_{n_x}$ & 0.318 & $1.71\times10^{-5}$ & $1.74\times10^{-5}$ & 1803 & 343 \\
        \hline
    \end{tabular}
\end{table}

\section{Conclusion} \label{sec:conclusion}
This paper presents a stochastic differential dynamic programming algorithm for partially observable trajectory optimization problems.
In the algorithm, the control update is computed by explicitly considering the coupled evolution of the nominal trajectory, the state-estimation process, and the feedback policy, without relying on the separation principle.
Constraints on thrust magnitude, terminal state, and covariance are handled using an augmented Lagrangian formulation.
For computational efficiency, the method computes state transition matrices for covariance propagation semi-analytically and uses automatic differentiation to evaluate model-dependent derivatives.

The numerical results demonstrate three distinct capabilities of the proposed approach.
In the light-dark domain problem, the method generates a navigation-aware solution that exploits an informative region while maintaining the path and thrust constraints and achieving terminal covariance performance close to the prescribed target.
In the Earth-to-Mars transfer, it produces uncertainty-robust solutions under nonlinear dynamics, nonlinear observations, and maneuver execution errors.
The periodic-orbit transfer in the Earth--Moon CR3BP shows that the method identifies solutions that exploit the coupling between trajectory design and orbit determination in a strongly nonlinear dynamical environment.
The results indicate that the proposed framework provides a practical way to incorporate navigation performance and feedback robustness into mission design problems under partial observability.

\appendix
\section*{Appendix: Computation of State Transition Matrices}
In practice, implementing derivatives of the dynamics up to second order is cumbersome because these derivatives depend on the specific structure of the dynamics.
To avoid manual derivations of the differential equations, AD can be applied directly to the discrete-time nonlinear dynamics $\bm{f}_{k,j}(\bm{x}_{k,j},\bm{u}_{k,j})$ to compute the STMs.
We use two Julia packages to compute the nominal-state STMs.
First, \texttt{DifferentialEquations.jl}~\cite{rackauckas2017differentialequations} is used to discretize the continuous dynamics $\bm{f}(\bm{x},\bm{u})$ with a prescribed numerical accuracy.
Then, \texttt{ForwardDiff.jl}~\cite{RevelsLubinPapamarkou2016} is applied to the discretized nonlinear dynamics $\bm{f}_{k,j}(\bm{x}_{k,j},\bm{u}_{k,j})$, providing an easy-to-use interface for forward-mode AD\@.
The second-order STMs can also be obtained through a nested approach by applying AD to the first-order derivatives.

The stage-to-stage STMs, $\Phi^1_k$ and $\Phi^2_k$, are then obtained by repeated application of the chain rule over the intermediate epochs.
The first-order STM is obtained by matrix multiplication:
\begin{align}
    \Phi^1_k = \Phi^1_{(t_{k,N_j(k)+1}, t_{k,N_j(k)})} \cdots \Phi^1_{(t_{k,1}, t_{k,0})}
\end{align}
where $\Phi^1_{(t_{k,j+1}, t_{k,j})}$ denotes the first-order STM from $t_{k,j}$ to $t_{k,j+1}$.
The second-order STM is obtained recursively in chronological order, for example, as
\begin{align}
    \Phi^{{i,ab}}_{(t_{k,j+2}, t_{k,j})}  = \Phi^{{i,\alpha}}_{(t_{k,j+2}, t_{k,j+1})} \Phi^{{\alpha,ab}}_{(t_{k,j+1}, t_{k,j})} + \Phi^{{i,\alpha \beta}}_{(t_{k,j+2}, t_{k,j+1})} \Phi^{{\alpha,a}}_{(t_{k,j+1}, t_{k,j})} \Phi^{{\beta,b}}_{(t_{k,j+1}, t_{k,j})}
\end{align}
where $\Phi^{{i,ab}}_{(t_{k,j+2}, t_{k,j})}$ represents the second-order STM from $t_{k,j}$ to $t_{k,j+2}$ and is defined as
\begin{align}
    \Phi^{{i,ab}}_{(t_{k,j+2}, t_{k,j})} = \frac{\partial^2 Z^i_{k,j+2}}{\partial Z^a_{k,j} \partial Z^b_{k,j}}
\end{align}

\section*{Code Availability}
The source code implementing the proposed algorithm and reproducing the numerical examples will be made publicly available upon publication at \url{https://github.com/mfuji074/PartiallyObservableSDDP.jl}.

\section*{Funding Sources}
This work was supported by the Japan Society for the Promotion of Science (JSPS) KAKENHI Grant-in-Aid for Early-Career Scientists, Grant Number JP24K17449.

\section*{Acknowledgments}
The authors used ChatGPT (OpenAI) during manuscript preparation to improve wording, readability, and clarity. All technical content and final edits were verified by the authors.

\bibliography{references}

\end{document}